\documentclass[10pt,journal,compsoc]{IEEEtran}
\ifCLASSOPTIONcompsoc
  \usepackage[nocompress]{cite}
\else
  \usepackage{cite}
\fi
\ifCLASSINFOpdf
   \usepackage[pdftex]{graphicx}
\else
\fi
\usepackage{amsmath}
\usepackage{algorithmic}
\usepackage{array}
\ifCLASSOPTIONcompsoc
  \usepackage[caption=false,font=footnotesize,labelfont=sf,textfont=sf]{subfig}
\else
  \usepackage[caption=false,font=footnotesize]{subfig}
\fi
\usepackage{fixltx2e}
\usepackage{stfloats}
\ifCLASSOPTIONcaptionsoff
  \usepackage[nomarkers]{endfloat}
 \let\MYoriglatexcaption\caption
 \renewcommand{\caption}[2][\relax]{\MYoriglatexcaption[#2]{#2}}
\fi
\usepackage{multirow}
\usepackage{amsfonts}
\usepackage{hhline}

\usepackage{booktabs}

\usepackage{makecell}
\setcellgapes{3pt}

\begin{document}
\title{ANAE: Learning Node Context Representation for Attributed Network Embedding}

\author{Keting~Cen, 
Huawei~Shen\IEEEauthorrefmark{1}\thanks{\IEEEauthorrefmark{1}Corresponding author}, 
Jinhua~Gao, 
Qi~Cao, 
Bingbing~Xu, 
Xueqi~Cheng~\IEEEmembership{Member,~IEEE,}
\IEEEcompsocitemizethanks{
  \IEEEcompsocthanksitem
            Keting Cen is with the CAS Key Laboratory of Network Data Science and Technology, Institute of Computing Technology (ICT), Chinese Academy of Sciences (CAS), Beijing, 100190, China.  
            \protect\\ E-mail:
            \protect cenketing@ict.ac.cn
  \IEEEcompsocthanksitem
            Huawei Shen is with the CAS Key Laboratory of Network Data Science and Technology, Institute of Computing Technology (ICT), Chinese Academy of Sciences (CAS), Beijing, 100190, China.
            \protect\\ E-mail:
            \protect shenhuawei@ict.ac.cn
            %\protect\\ \IEEEauthorrefmark{1}Corresponding author 
  \IEEEcompsocthanksitem
            Jinhua Gao, Qi Cao, Bingbing Xu and Xueqi Cheng are with the CAS Key Laboratory of Network Data Science and Technology, Institute of Computing Technology (ICT), Chinese Academy of Sciences (CAS), Beijing, 100190, China.
            \protect\\ E-mail:
            \protect \{gaojinhua,caoqi,xubingbing,cxq\}@ict.ac.cn
}
}
\IEEEtitleabstractindextext{%
\begin{abstract}
Attributed network embedding aims to learn low-dimensional node representations from both network structure and node attributes. Existing methods can be categorized into two groups: (1) the first group learns two separated node representations from network structure and node attribute respectively and concatenates them together; (2) the other group obtains node representations by translating node attributes into network structure or vice versa. However, both groups have their drawbacks. The first group neglects the correlation between network structure and node attributes, while the second group assumes strong dependence between these two types of information. In this paper, we address attributed network embedding from a novel perspective, i.e., learning node context representation for each node via modeling its attributed local subgraph. To achieve this goal, we propose a novel attributed network auto-encoder framework, namely ANAE. For a target node, ANAE first aggregates the attribute information from its attributed local subgraph, obtaining its low-dimensional representation. Next, ANAE diffuses the representation of the target node to nodes in its local subgraph to reconstruct their attributes. 
Such an encoder-decoder framework allows the learned representations to better preserve the context information manifested in both network structure and node attributes, thus having high capacity to learn good node representations for attributed network. Extensive experimental results on real-world datasets demonstrate that the proposed framework outperforms the state-of-the-art approaches at the tasks of link prediction and node classification.
%Attributed network embedding aims to learn low-dimensional node representations from both network structure and node attributes. Existing methods can be categorized into two groups: (1) the first group learns two separate node representations from network structure and node attributes respectively and concatenates them together; (2) the other group obtains node representations by translating node attributes into network structure or vice versa. However, both groups have their drawbacks. The first group neglects the correlation between network structure and node attributes, while the second group assumes strong dependences between these two types of information. In this paper, we address attributed network embedding from a novel perspective, i.e., learning context representation for each node via modeling its attributed local subgraph. To achieve this goal, we propose a novel attributed network auto-encoder framework (ANAE). For a target node, ANAE first aggregates the attributes from its attributed local subgraph, obtaining its low-dimensional representation. Next, ANAE diffuses the representation to nodes in its local subgraph to reconstruct their attributes. 
%Such an encoder-decoder framework better preserves the context information in both network structure and node attributes, thus getting better representations. Extensive experimental results on real-world datasets demonstrate that ANAE outperforms the state-of-the-art approaches.
\end{abstract}

\begin{IEEEkeywords}
Attributed Network Embedding, Attributed Network Auto-encoder, Attributed local subgraph, Node Context Representation, Link Prediction, Node Classification.
\end{IEEEkeywords}}

\maketitle

\IEEEdisplaynontitleabstractindextext

% \fi
\IEEEpeerreviewmaketitle

%\IEEEraisesectionheading{\section{Introduction}\label{sec:introduction}}

\IEEEraisesectionheading{\section{Introduction}\label{sec:introduction}}
%\section{Introduction}
%Networks are ubiquitous in nature and society, including social networks, information networks, biological networks and various technological networks. The connections among nodes of network form a major obstacle for many data mining tasks dealing with networks. For example, when computing the PageRank score of nodes in a network, we need a number of iterations to extract relevant information from the structure of network~\cite{page1999pagerank}; for graph-based semi-supervised learning, it is necessary and usually valuable to consider the diffusion of node attributes or labels over network, leveraging graph structure as regularization. ~\cite{yang2016revisiting, kipf2017semi}. To combat the challenge of mining network data, researchers attempt to learn a low-dimensional representation for each node in network, which can capture and preserve the network structure~\cite{cui2018survey}. With the learned representations of nodes, many downstream mining and prediction tasks on networks, e.g., node classification and link prediction, can be easily addressed using standard machine learning tools. The problem of learning representations of nodes is usually called \emph{network embedding} in the community of data mining.

\IEEEPARstart{N}etworks are ubiquitous in nature and society, including social networks, information networks, biological networks and various technological networks. The complex structure of networks poses big challenge for data mining tasks dealing with networks. To combat this challenge, researchers resort to network embedding, i.e., learning low-dimensional representation for each node to capture and preserve network structure~\cite{cui2018survey,grover2016node2vec,perozzi2014deepwalk}. With the learned representations of nodes, many downstream mining and prediction tasks on networks, e.g., node classification and link prediction, can be easily addressed using standard machine learning tools. 

Many network embedding methods have been proposed and successfully applied to node classification and link prediction~\cite{cui2018survey, tang2015line, ou2016asymmetric,cao2015grarep}. 
These methods leverage only structural information of nodes, i.e., structural proximity and structural similarity, to learn node representations.
%These methods learn representations of nodes leveraging structural proximity or structural similarity among nodes. 
However, in many real world networks, nodes are usually associated with rich attributes, e.g., content of articles in citation network ~\cite{le2014probabilistic} and user profile in social networks ~\cite{qi2012exploring,huang2018exploring}. This motivates researchers to study the problem of attributed network embedding. 

Attributed network embedding aims to learn a low-dimensional representation for each node by simultaneously considering the information manifested in both network structure and node attributes~\cite{liao2018attributed,huang2017accelerated,huang2017label,Shah:2019:GAN:3308558.3313640,bandyopadhyay2019outlier}. Existing methods for attributed network embedding mainly fall into two paradigms. Methods in the first paradigm learn two separated representations for each node according to network structure and node attributes respectively, and then concatenate them into a single representation ~\cite{liao2018attributed, gao2018deep,zhang2018anrl}. Methods in the other paradigm attempt to directly obtain a single representation for each node by translating node attributes into network structure or vice versa~\cite{yang2015network, liu2018content}, which are later referred to as translation models. However, methods in both paradigms have their drawbacks. Methods in the first paradigm neglect the correlation between these two types of information, while the second paradigm assumes strong dependence between node attributes and network structure. Thus we are still lack of an effective method for attributed network embedding. 

%Recently some graph neural network based models have been proposed to learn node representation, i.e., GAE~\cite{kipf2016variational}, DGI\cite{velickovic2018deep}. GAE learn node representation by employing GCN layer~\cite{kipf2017semi} to reconstruct network structure only. Thus this method suffer with loss of attribute information. DGI also adopt GCN to learn node representation but aim at maximizing mutual information between patch representations and corresponding high-level summaries of graphs\cite{velickovic2018deep}. The difficulty of calculation of mutual information makes the model much complicated.

In this paper, we propose a novel perspective to address attributed network embedding. Unlike previous methods, we attempt to learn the representation of each node by modeling its local context, i.e., attributed local subgraph, which is defined as the subgraph centered at the target node together with associated node attributes.
From our perspective, good representation for each node in attributed networks should be generated from the information manifested in its local context, as well as be capable of recovering its local context.
%the representation of each node in attributed networks should well capture both structural information and attribute information manifested in its attributed local subgraph.
%This perspective transfroms the problem of learning node representations into the problem of modeling the context information manifested in both network structure and node attributes.
 Motivated by this perspective, we propose a novel attributed network auto-encoder framework, namely ANAE, for attributed network embedding. ANAE consists of graph context encoder and graph context decoder. In graph context encoder, target node aggregates the attribute information from nodes in its local subgraph to generate its own representation. In graph context decoder, each node diffuses its representation to nodes in its local subgraph to help reconstruct their attributes.
Our proposed framework allows the representation of each node to preserve useful context information manifested in its attributed local subgraph as much as possible via an encoder-decoder framework, having high capacity to learn good node representations for attributed networks. 
%Our proposed framework guarantees the consistency between the representation of each node and the context information manifested in its attributed local subgraph via an encoder-decoder process, 
 %Our proposed framework generates a node's representation by capturing both the structural information and attribute information manifested in its attributed local subgraph, 

The main contributions are summarized as follows:
\begin{itemize}
\item 
We propose a novel perspective for attributed network embedding, i.e., transforming the problem of learning node representations into the problem of modeling the attributed local subgraph of node. From our perspective, the representations of each node should preserve the context information manifested in its attributed local subgraph.
%A node's representation should well capture the context information manifested in its attributed local subgraph.
%3We transfroms the problem of learning node representations into the problem of modeling the attributed local subgraph, which naturally capture network structure, node attributes and their interaction simultaneously.
\item 
Motivated by our perspective, we propose a novel, flexible auto-encoder framework, ANAE, for attributed network embedding. Moreover, we also give an implementation of the framework with graph attention network serving as building blocks of encoders and decoders.
%An implementation is also provided with graph attention layer serving as encoders and decoders.
%which encodes attributed local subgraph of target node into hidden representation and reconstruct the attributed local subgraph at decoder. 
%Motivated by our perspective, we propose a novel and flexible auto-encoder framework, ANAE, for attributed network embedding. An implementation is also provided with graph neural networks serving as encoders and decoders.
\item Extensive experiments are conducted on real-world attributed networks at the tasks of node classification and link prediction. The results demonstrate the superior performance of our proposed model over baselines. Moreover, ablation analysis is provided to illustrate the effectiveness of different components of ANAE.
%We perform extensive experiments on real-world attributed network to corroborate the efficacy of ANAE of several downstream tasks(link prediction and node classification) and give detailed analysis on the results. Experimental results on real-world datasets demonstrate that our proposed method outperforms the state-of-the-art network embedding approaches at both tasks. Moreover we compare ANAE with some of its variants to corroborate the efficacy of each components of ANAE through experiments.

\end{itemize}

The rest of this paper is organized as follows: Section \ref{sec2} briefly reviews related works. The problem formulation and framework overview of ANAE are introduced in Section \ref{sec3}. In Section \ref{sec4} we present our implementation of ANAE. The experiments and results are presented in Section \ref{sec5}, while detailed analysis of the performance of ANAE is provided in Section \ref{sec6}. Finally, Section \ref{sec7} concludes the paper and describes the future work.

\section{Related Work}\label{sec2}
Our proposed framework works in an encoder-decoder manner to learn better embeddings for attributed networks. To simultaneously capture the network structures and node attributes manifested in attributed local subgraph, graph convolutional network is adopted in both encoder layer and decoder layer. In this section, we provide a brief introduction of related works in network embedding and graph convolutional network.
\subsection{Plain Network Embedding}
Network embedding technology, which aims to learn low-dimensional embedding for nodes in network, actually evolves from the problem of dimension reduction of graph data~\cite{cui2018survey}. Some early works first leverage feature similarity to build an affinity graph, and then treat eigenvectors as network representations, such as LLE~\cite{roweis2000nonlinear} and Isomap~\cite{tenenbaum2000global}. Recently, more network embedding methods leveraging the structural proximity or structural similarity among nodes have been proposed. Structural proximity based methods try to preserve different orders of proximities among nodes when learning node embeddings, varying from first-order proximity~\cite{man2016predict}, second order proximity~\cite{tang2015line,wang2018graphgan} to high order proximity~\cite{cao2015grarep,wang2017community,ou2016asymmetric,feng2018representation,qiu2018network,ma2018multi}. Moreover, some deep models~\cite{cao2016deep, wang2016structural} have been proposed to account for more complex structural properties.  Structural similarity based approaches~\cite{henderson2012rolx, ribeiro2017struc2vec, donnat2018} take into account structural roles of nodes, restricting nodes with similar structural roles to possess similar representations. 
%Besides, some approaches are proposed to capture structural similarity of nodes~\cite{henderson2012rolx, ribeiro2017struc2vec, donnat2018}. Since the network structure is complex, shallow models cannot capture the highly non-linear network structure, some deep models~\cite{cao2016deep, wang2016structural} are proposed to solve this problem. 
\subsection{Attributed Network Embedding}
All the above mentioned approaches are limited to deal with plain networks. Besides structural properties, nodes in real world networks are usually associated with rich labels and attributes, which contributes to the problem of attributed network embedding. Recently, much efforts have been made to gain insights from attributed networks~\cite{yang2015network,liao2018attributed,le2014probabilistic,zhu2007combining,qi2012exploring,yang2018binarized,li2017attributed}. Some approaches~\cite{tu2016max, huang2017label, huang2017accelerated} simply take the label information into consideration, while others utilize more detailed attribute information. The key point of attributed network embedding lies in simultaneously capturing node attributes, network structure and their relationship into hidden representations. TADW~\cite{yang2015network} proposes to obtain node embeddings by decomposing the adjacency matrix, with the attribute matrix being fixed as a factor. DANE~\cite{gao2018deep} leverages two separated auto-encoders to learn structural representations and attributed representations of nodes respectively and concatenates them as final representations with consistent and complementary regularization in hidden layer. Inspired by machine translation, STNE~\cite{liu2018content} obtains the node representations by translating the sequence of node attributes to the sequence of node identities.
GAE ~\cite{kipf2016variational} adopts GCN layer ~\cite{kipf2017semi} to encode both network structure and node attributes into representations and refines them by reconstructing network structure. In fact, GAE can be considered as a special case of ANAE, but it fails to capture the attribute information in decoder.
%Recently, GAE~\cite{kipf2016variational} a special case of our ANAE framework, which uses GCN layer~\cite{kipf2017semi} as encoder but only aim at reconstructing network structure at decoder. Representations learned by GAE will lose useful information in node attributes.
%Attributed network embedding takes node attributes into consideration when learning embedding has attracted a lot attention~\cite{yang2015network,liao2018attributed,le2014probabilistic,zhu2007combining,qi2012exploring}. Label information plays an important role in attribute network embedding~\cite{tu2016max, huang2017label, huang2017accelerated}. Recent method DANE~\cite{gao2018deep} learns separated representations for network structure and node attributes with consistent or complementary constraint in hidden layer. TADW~\cite{yang2015network} learns embedding to infer network structure with node attributes as regularization.

%Different from the above-mentioned algorithms that learn embeddings on homogeneous networks, some works~\cite{chang2015heterogeneous,dong2017metapath2vec,shi2016survey,cen2019representation,wang2018shine} also investigate network embedding on heterogeneous networks that have different types of nodes and links.

\subsection{Graph Convolutional Network}
%kipf2017semi, velickovic2018graph
Graph convolutional networks which generalize convolution neural networks to non-Euclidean domains~\cite{bronstein2017geometric} have gained remarkable success in various tasks. Existing methods can be roughly categorized into two kinds, i.e., spectral methods~\cite{bruna2014spectral,defferrard2016convolutional,DBLP:conf/ijcai/XuSCCC19} and spatial methods~\cite{bronstein2017geometric,DBLP:conf/iclr/XuSCQC19}. Spectral methods define the convolution on spectral domain, which first transform the signal into spectral domain and apply filters on it~\cite{bruna2014spectral}. Spatial methods view graph convolution as ``patch operator'', which constructs a new feature vector from its neighboring nodes~\cite{monti2017geometric}. GCN introduced by Kipf et al.~\cite{kipf2017semi} defines convolution operator as the weighted sum of neighboring nodes' features and weights are defined by normalized edge weights. As the weights in GCN are determined by network structure only, Velickovic et al.~\cite{velickovic2018graph} proposes Graph Attention Network~(GAT) to learn the weights by structure masked self-attention. 
%GAT is a supervised model aims to map input data into their labels, while our ANAE learns representation for each node by reconstruct their local attributed subgraph in an unsupervised manner.
Moreover, to extend GCN to large-scale networks, GraphSAGE~\cite{hamilton2017inductive} learns a function that generates embeddings by sampling and aggregating features from a node's local neighborhood, which could learn inductive node embedding for large graph. Some recent works ~\cite{chen2018stochastic,DBLP:conf/iclr/ChenMX18,ying2018graph,jacob2014learning} further optimize the sampling strategy so that GCN can be better applied to large-scale networks. 

\begin{figure}[!t]
\centering
\subfloat[Independently model attributes and structure]{\includegraphics[width=0.5\textwidth]{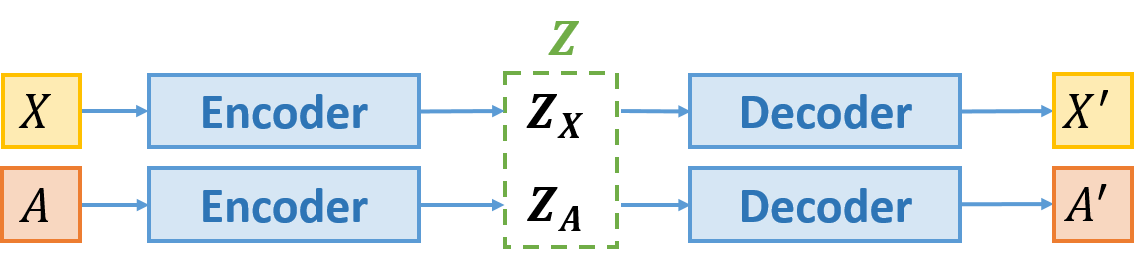}%
\label{framepict_1}}
\hfil
\subfloat[Translation model]{\includegraphics[width=0.5\textwidth]{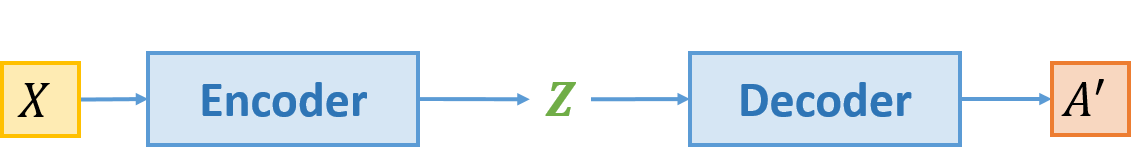}%
\label{framepict_2}}
\hfil
\subfloat[GAE]{\includegraphics[width=0.5\textwidth]{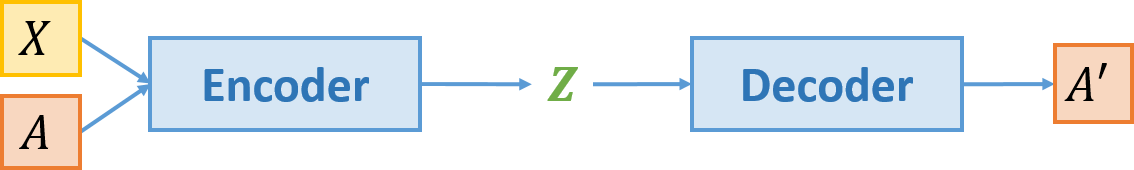}%
\label{framepict_3}}
\hfil
\subfloat[ANAE]{\includegraphics[width=0.5\textwidth]{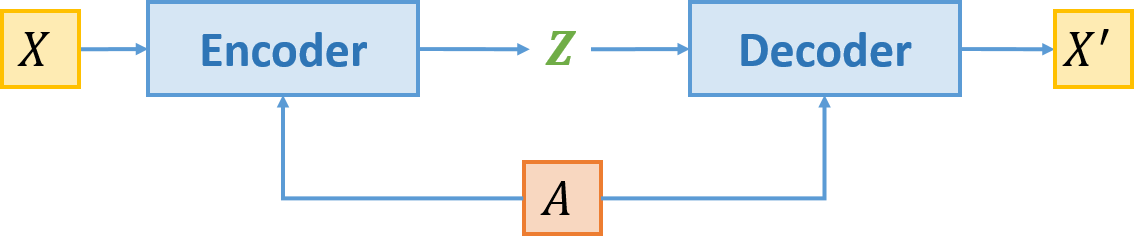}%
\label{framepict_3}}
\caption{Comparing our architecture with previous frameworks. $X$ denotes the node attributes, $X'$ represents the reconstructed node attributes. Network structure and reconstructed network structure are given as $A$ and $A'$ respectively. $Z$ is node representations. Specifically in (a) $Z$ is obtained by concatenating $Z_X$ learned from attributes and $Z_A$ learned from network structure.}
\label{framepict}
\end{figure}

%\begin{figure}
%\centering
%\begin{tabular}{l}
%	\begin{minipage}[c]{0.5\textwidth}
%		\includegraphics[width=\textwidth]{pict/seperate-simple.png}	
%		\caption*{(a) Independently model attributes and structure}
%	\end{minipage}
%\end{tabular}
%\begin{tabular}{l}
%	\begin{minipage}[c]{0.5\textwidth}
%		\includegraphics[width=\textwidth]{pict/translate-simple.png}	
%		\caption*{(b) Translation model}
%	\end{minipage}
%\end{tabular}
%\begin{tabular}{l}
%	\begin{minipage}[c]{0.5\textwidth}
%		%\includegraphics[width=\textwidth]{pict/model-structure.png}
%		%\includegraphics[width=\textwidth]{pict/model-structure-new.png}
%		\includegraphics[width=\textwidth]{pict/GAE-simple.png}
%		\caption*{(c) GAE}
%	\end{minipage}
%\end{tabular}
%\begin{tabular}{l}
%	\begin{minipage}[c]{0.5\textwidth}
%		%\includegraphics[width=\textwidth]{pict/model-structure.png}
%		%\includegraphics[width=\textwidth]{pict/model-structure-new.png}
%		\includegraphics[width=\textwidth]{pict/ANAE-simple.png}
%		\caption*{(d) ANAE}
%	\end{minipage}
%\end{tabular}
%\caption {Comparing our architecture with previous frameworks. $X$ denotes the node attributes, $X'$ represents the reconstructed node attributes. Network structure and reconstructed network structure are given as $A$ and $A'$ respectively. $Z$ is node representations. Specifically in (a) $Z$ is obtained by concatenating $Z_X$ learned from attributes and $Z_A$ learned from network structure.}
%\label{framepict}
%\end{figure}
%%\fi

\section{Problem Formulation and ANAE Framework}\label{sec3}
In this section, we first provide the problem formulation of attributed network embedding. Then we introduce our proposed ANAE framework, and illustrate its advantages by comparing with existing attributed network embedding methods.

\subsection{Problem Formulation}
We denote a network as $G = \{V, E\}$, where $V=\{v_i\}_{i=1}^{n}$ denotes the node set with size $n$, and $E \subseteq V \times V$ denotes the edge set. The network is represented by an adjacency matrix A, where $A_{ij}$ is the weight of edge $E_{ij}$ and $A_{ij} = 0$ otherwise.
Attributes of nodes in the network are represented by an attribute matrix $X \in \mathbb{R}^{n\times d}$, where $d$ is the dimension of node attributes. $X_i$ is the $i$-th row of $X$ representing attribute vector of node $v_i$.

Attributed network embedding aims to learn low-dimensional representations $Z \in \mathbb{R}^{n \times k}$ from adjacency matrix $A$ and attribute matrix $X$, such that the learned representations can preserve information manifested in both network structure and node attributes. 
%Formally, the purpose of attributed network embedding is to learn a function $f(A, X) \rightarrow Z$, where $Z \in \mathbb{R}^{n \times k}$ denotes the matrix of embedding that capture both network structure and node attributes, and $k$ is the dimension of embedding. 
We summarize the notations and their descriptions in Table \ref{notation}.

\begin{table}[h]
\centering
\makegapedcells
\caption{Notations and Terms}
\begin{tabular}{|c|l|}
\hline
%\toprule
Notations & Descriptions \\
\hhline{==}
$A$ & adjacency matrix \\
$A^{\prime}$ & reconstructed adjacency matrix \\
$X$ & node attribute matrix \\
$X^{\prime}$ & reconstructed node attribute matrix \\
$Z$ & node embedding matrix \\
$F(A,H)$ & graph encoder layer \\
$F^{\prime}(A,H)$ & graph decoder layer \\
$H^{t}$ & hidden representation matrix in $t$-th encoder layer \\
$H^{\prime t}$ & hidden representation matrix in $t$-th decoder layer \\
$h^{t}_{i}$ & hidden representation of node $i$ in $t$-th encoder layer \\
$h^{t \prime}_{i}$ & hidden representation of node $i$ in $t$-th decoder layer \\
$\mathcal{N}_i$ & neighbor nodes of node $i$\\
$W_t$ & the weight matrix in $t$-th encoder layer \\
$W^{\prime}_t$ & the weight matrix in $t$-th encoder layer \\
$\alpha_{ij}$ & the aggregation weights between node $i$ and $j$ \\
$\vec{a}$ & the weight vector in attention mechanism\\
\hline
%\bottomrule
\end{tabular}
\label{notation}
\end{table}

\subsection{ANAE Framework}
The key challenge to attributed network embedding lies in capturing the information manifested in both network structure and node attributes. In this paper, we propose a novel perspective to address attributed network embedding, i.e., learning node representation by modeling its local context. We define the local context of a node to be its attributed local subgraph, i.e., the subgraph centered at the target node together with associated node attributes. 
To preserve effiective information from local context as much as possible, good representation for each node in attributed networks should be generated from the information manifested in its local context, as well as be capable of recovering the local context. 

To achieve this goal, we propose a novel encoder-decoder framework named ANAE. As shown in Figure \ref{framepict}(d), ANAE consists of a graph context encoder and a graph context decoder. Both graph context encoder and graph context decoder take the network structure~($A$) as input to specify the local subgraph of each node. For each node, graph context encoder aggregates all attribute information in its local subgraph to generate its representation, while graph context decoder refines its aggregated representation by reconstructing the attributes of nodes in its local subgraph. Such an encoder-decoder framework allows the learned representation of each node to preserve the context information manifested in its attributed local subgraph, and possesses high capacity to learn good context representation for nodes in attributed networks.
%We illustrate the superiority of our proposed framework by comparing to several representative previous models (Figure \ref{framepict}). Figure \ref{framepict} (a) shows models that learn two separated representations for each node according to network structure and node attributes respectively, and then concatenate them into a single representation ~\cite{liao2018attributed, gao2018deep,zhang2018anrl}, while Figure \ref{framepict} (b) presents the architecture of translation models ~\cite{yang2015network, liu2018content}. The former models neglect the dependence between network structure and node attributes, while the latter models assume too strong dependence between these two. Figure \ref{framepict} (c) depicts the architecture of GAE ~\cite{kipf2016variational}. GAE encodes network structure and node attributes to generate node representations and refines them by reconstructing network structure in the decoder, ignoring the attribute information. Compared to GAE, ANAE takes network structure into graph context decoder to specify the local subgraph of each node and aims to recover the attribute information, better capturing the context information manifested in the local context of each node.

We illustrate the advantage of our proposed ANAE framework by comparing it to several representative models in Figure \ref{framepict}. A typical solution (Figure \ref{framepict}(a)) is to learn two separated representations for each node according to network structure and node attributes respectively, and then concatenate them into a single representation~\cite{liao2018attributed, gao2018deep,zhang2018anrl}. However, this framework neglects the correlation between network structure and node attributes. Translation models (Figure \ref{framepict}(b)) generate the representations by transforming one type of information into the other, which assume too strong dependence between these two types of information. GAE (Figure \ref{framepict}(c)) aims to recover the network structure from hidden representations generated from node attributes and network structure, ignoring the attribute information in decoder. Moreover, both translation models and GAE try to recover one type of information in the decoder with the other type serving as input in the encoder, assuming strong correlation between network structure and node attributes. Unlike these models, ANAE learns the representation of each node via modeling its local context, i.e., its local subgraph and associated node attributes. For each node, ANAE specifies its local subgraph according to the network structure, and obtains its representation with attribute information in its local subgraph serving as input for encoder and output for decoder respectively. Therefore, our proposed ANAE framework can better capture the structural information and attribute information manifested in the local context of each node.

\begin{figure*}[!t]
\centering
\subfloat[Architecture of ANAE framework]{\includegraphics[width=\textwidth]{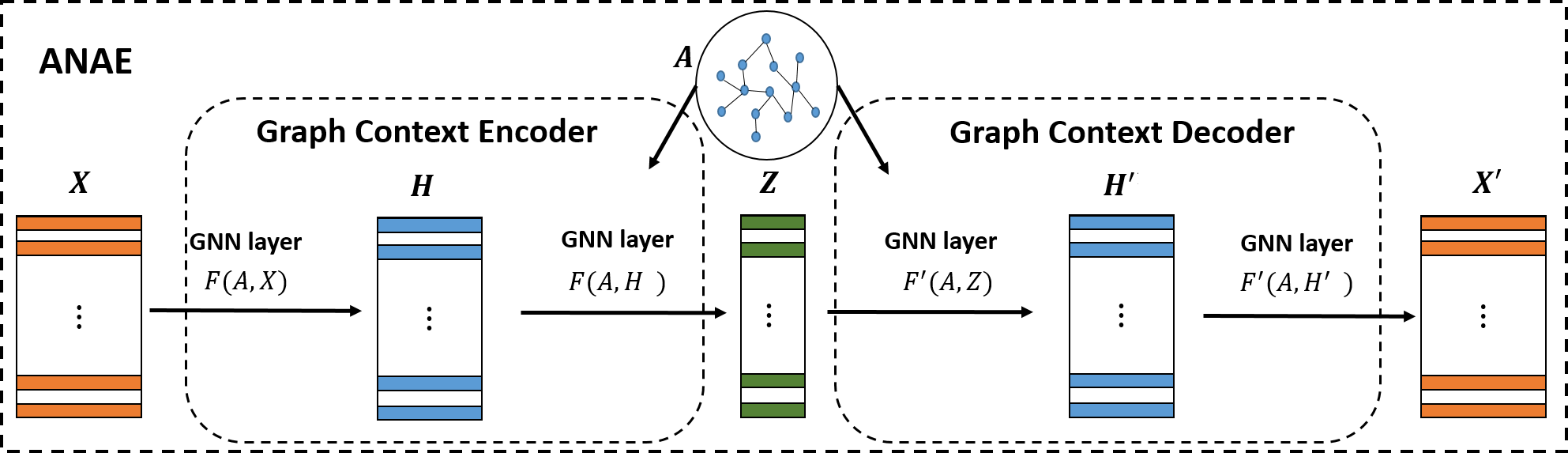}%
\label{modeldetail_1}}
\hfil
\subfloat[Single layer of graph context encoder]{\includegraphics[width=0.5\textwidth]{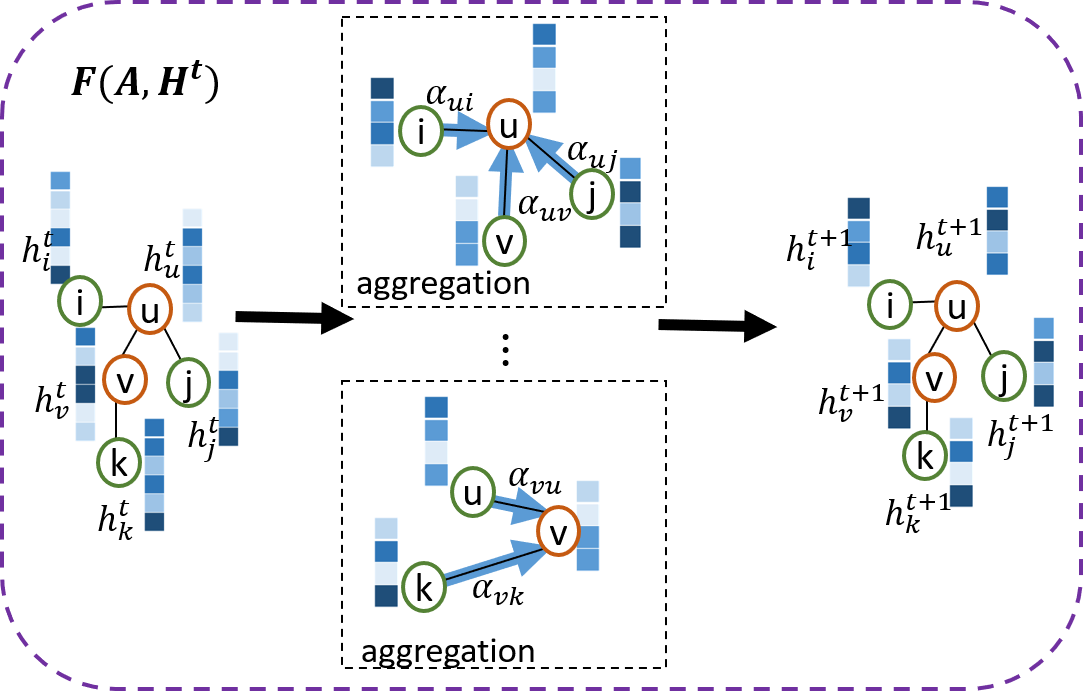}%
\label{modeldetail_2}}
\subfloat[Single layer of graph context decoder]{\includegraphics[width=0.5\textwidth]{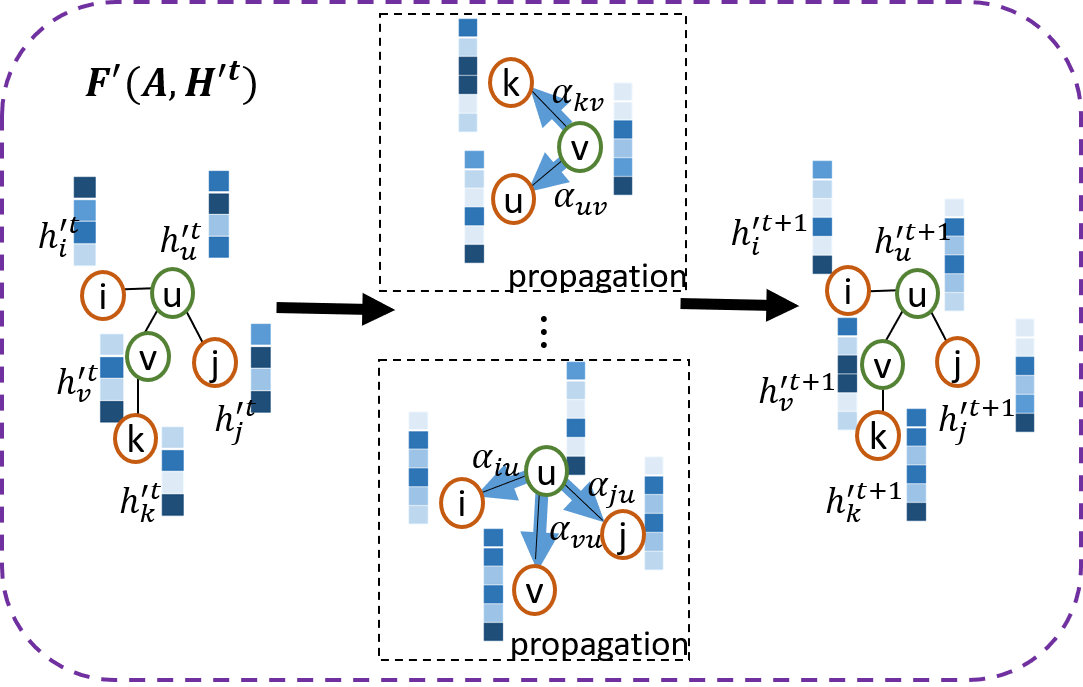}%
\label{modeldetail_3}}
\caption{Visual illustration of our implementation of ANAE. (a)~is the overall of our implementation, which encodes network structure $A$ and node attributes $X$ into representation $Z$ using graph context encoder and then recover node attribute $X$ from $Z$ via graph context decoder. $H^{t}$ represents the output of $t$-th layer in graph context encoder and $h_i^t$ is the i-th row of $H^{t}$ represents representation of node i, while $H^{\prime t}$ is the output of $t$-th layer in graph context decoder. $\alpha_{uv}$ is the aggregation~(propagation) weight in (b)~((c)) and arrow represents the direction of feature propagation. (b)~shows the process of a single layer of graph context encoder with nodes $u$ and $v$ as targets, $\alpha_{uv}$ is the aggregation weight. (c)~illustrates how nodes $u$ and $v$ propagate their representations to help neighboring nodes reconstruct their attributes in a single layer of graph context decoder.}
\label{modeldetail}
\end{figure*}

\section{An Implementation of ANAE}\label{sec4}

In this section, we provide an implementation of our ANAE framework, with graph neural networks (GNN) serving as building blocks for graph context encoder and graph context decoder, shown in Figure \ref{modeldetail}(a). Graph context encoder generates hidden representation $Z$ with attribute matrix $X$ serving as input, while graph context decoder tries to reconstruct such attribute matrix $X$ from hidden representation $Z$. The network structure $A$ is taken as input to specify the local subgraph of each node for both graph context encoder and graph context decoder. Specifically, we stack two GNN layers as graph context encoder while stack another two as graph context decoder. Both graph context encoder and graph context decoder characterize the diffusion of attribute information over network A. 
%Note that, various graph neural networks can be adopted to build graph context encoder and graph context decoder, e.g., GAT and GCN.

A single layer of graph context encoder and graph context decoder are shown in Figure \ref{modeldetail}(b) and Figure \ref{modeldetail}(c) respectively. Graph context encoder learns a sole embedding for the target node by aggregating the attribute information from nodes in its attributed local subgraph. In graph context decoder, each node propagates its representation to nodes in its local subgraph to help them reconstruct their attributes. The detailed implementation of graph context encoder and graph context decoder is described as follows.

%\subsubsection{\textbf{Graph Context Encoder}}
\subsection{Graph Context Encoder}

%In this section, we introduce the detailed implementation of graph context encoder.
%In this paper, we attempt to learn representation for each node that preserve its context, i.e., the attributed local subgraph. A node's attributed local subgraph is defined as the subgraph centered at the target node together with associated node attributes, which can be well captured by stacking sufficient number of graph convolution layers. Thus 
Our graph context encoder consists of a stack of single encoder layers, each of which aggregates the attribute information from the neighboring nodes of a target node. By stacking multiple encoder layers, graph context encoder is able to aggregate the attribute information from the multi-hop ego-network of the target node, which is taken as the target node's attributed local subgraph. 

%design of a single encoder layer $F(A,X)$, which takes the adjacency matrix together with the attribute matrix as input and encodes them into a low-dimensional representation for each node simultaneously. 
%In this paper we adopt graph attention network layer~\cite{velickovic2018graph} as a single graph context encoder layer, as the operators of which can be broken down into feature transformation and feature diffusion over graph. A main advantage of using GAT as encoder layer is that the aggregation weights are learned by node attributes and network structure, thus two unrelated nodes will have low aggregation weight even though them are directly connected In a word. 
%Encoder layer first aggregates the attribute information along the network structure, embedding of nodes become similar with their neighboring nodes. In a word such encoder layer directly encode the attributed subgraph into the representation of target node, and thus the representation contains rich information and get better result in downstream tasks.

A single encoder layer is formalized as follows:
\begin{equation}
h^{t+1}_{i} = \sigma(\sum_{j \in \mathcal{N}_i} \alpha_{ij} W_{t}h^{t}_{j}),
\end{equation} 
where $h^{t}_i \in \mathbb{R}^{f_t}$ is the hidden representation of node $i$ in the $t$-th layer and $f_t$ denotes its dimension, $\mathcal{N}_i$ is the set of node $i$'s neighbors including node $i$ itself in our experiments, $\alpha_{ij}$ is the aggregation weight and measures how important node $j$ is to node $i$, and $\sigma$ represents the nonlinear activation function. The linear transformation parameterized by a weight matrix $W_{t} \in \mathbb{R}^{f_{t}^{\prime} \times f_{t}}$ is applied on every node to extract effective high-level features from inputs, where $f_{t}^{\prime}$ is the dimension of output representation. The key of designing a single encoder layer lies in the definition of aggregation weight $\alpha_{ij}$, we which will be introduced in Section\ref{am}.

\subsection{Graph Context Decoder}
%Graph context encoder learns context representation for each node, while in graph context decoder we aim to reconstruct their attributed local subgraphs from their hidden representations. In other word, we attempt to recover both attributes of center node and attributes of nodes in its k-hop ego-network from its hidden representation. Our graph context decoder is significantly different from decoder in conventional auto-encoder which only aim to recover input of a sample from its hidden representation.
Graph Context decoder aims to reconstruct the attributed local subgraph of each node from the hidden representation, which is obtained by graph context encoder. 
%In other word, we attempt to recover both attributes of target node and attributes of nodes in its k-hop ego-network from its hidden representation. 
It is worth noting that our graph context decoder is significantly different from the decoder in conventional auto-encoder framework, which only aims to recover the attribute of target node itself from its hidden representation.
%To recover attributes we aim to refine the identity information of a target node from its hidden representation, while to recover attributed local subgraph 

To recover the attributed local subgraph of each node, graph context decoder incorporates the network structure to specify the local subgraph for each node and propagates hidden representation of a target node to nodes in its local subgraph. 
%To achieve this goal, we add feature transform to refine useful information and feature propagation in each decoder layer. 
In this layer each node propagates its hidden representation to its neighbor nodes and help them reconstruct their attributes, which implies that hidden representation of a target node contains sufficient information of its attributed local subgraph. 
%In fact, all nodes propagating to their neighbors are identical to all nodes aggregating from their neighbors from the global view of the whole network. 
Taking Figure \ref{modeldetail}(c) as example, node $v$ propagates its representation to neighbors $u$, $k$ with attention weight $\alpha_{kv}$ and $\alpha_{uv}$. From the view of node $k$, this operator is equal to aggregate representation from node $v$ with attention weight $\alpha_{kv}$. 
%The overall architecture for graph context decoder is shown right side of Figure \ref{modeldetail}(a) and a single layer of graph context decoder is shown in Figure \ref{modeldetail}(c). 
Thus we adopt the following architecture as graph context encoder and formalize it as follows:
\begin{equation}
h^{'t+1}_{i} = \sigma(\sum_{j \in \mathcal{N}_i} \alpha_{ij} W^{'}_{t}h^{t}_{j}).
\end{equation} 

Note that, the encoder layer compresses node representations into lower dimension, while decoder layer uncompresses them. In our experiments the hidden units in graph context decoder are symmetric to graph context encoder. 
%For example, if we stack two layer graph attention layer in graph context encoder and let dimension of first layer output be $d_1$ and the second be $d_2$, then the dimension of output in first layer and second layer of graph context decoder are $d_2$ and $d_1$ respectively. 

%\subsubsection{\textbf{Aggregation mechanism}}\label{am}
\subsection{Aggregation mechanism}\label{am}

Aggregation weights measure how important neighboring nodes are to the target node, 
%edge weights natural fit this but will fail in some scenarios, e.g., network lacking edge weights or edge weights irrelevant to similarity. To handle these drawbacks, 
which are crucial to the performance of our proposed framework. In both graph context encoder and graph context decoder, the local subgraphs of nodes have already be specified by input network structure~($A$) and edge weights may not reflect the similarity of nodes accurately. Thus we adopt shared attention mechanism to learn the aggregation weight between two given nodes according to their attributes. Note that, various aggregation mechanism can be adopted, e.g., attention mechanism and edge weights. We use the same attention mechanism as GAT~\cite{velickovic2018graph} in this paper. Detailed analysis of the performance of different choices can be found in Section \ref{sec6.1}. Intuitively, this mechanism is a single layer neural network, which is parameterized by a weight vector $\vec{a} \in \mathbb{R}^{2f^{\prime}}$, following with a nonlinear activation. For each node $i$, we only compute $\alpha_{ij}$ for node $j \in \mathcal{N}_i$, where $\mathcal{N}_i$ denotes the neighbors of node i~\cite{velickovic2018graph}. The attention mechanism can be expressed as:
%We also propose to measure the aggregation weight from the attribute view. And attention mechanism provides an effective way to achieve this goal. In our experiments, we use the same attention mechanism named self-attention as GAT~\cite{velickovic2018graph}. Self attention mechanism can be implemented as a single-layer feedforward neural network that is parametrized by a weight vector $\vec{a} \in \mathbb{R}^{2d^{\prime}}$($d^{\prime}$ is the dimension of input vector), following with a nonlinear activation. The attention mechanism can be expressed as:
\begin{equation}
\alpha_{ij} = \frac{\mathrm{exp}(\sigma^{\prime}(\vec{a}^{\mathrm{T}}[W\vec{h}_i||W\vec{h}_j]))}{\sum_{k\in \mathcal{N}_i}\mathrm{exp}(\sigma^{\prime}(\vec{a}^{\mathrm{T}}[W\vec{h}_i||W\vec{h}_k]))} ,
\end{equation}
where $\cdot^{\mathrm{T}}$ denotes matrix transposition and $||$ represents concatenation operation. In our experiments, we adopt LeakyReLU ~(with negative input slope set to 0.2)~\cite{velickovic2018graph} as nonlinear activation $\sigma^{\prime}$. We also employ multi-head attention to stabilize the learning process of self-attention and capture multiple types of relationships between nodes. In our experiments, we concatenate the representations learned by different heads in each hidden layer, and average them on the final layer of the graph context encoder. 

Attention based aggregation mechanism can capture both the structural proximity and the attribute proximity between pairs of nodes, allowing better modeling the attributed local subgraph. 

%\subsubsection{\textbf{Loss Function}}
\subsection{Loss Function}
The context of a object reflects the characteristics of it, many models (e.g. CBOW, Skip-Gram\cite{mikolov2013distributed}) have verified this hypothesis. We propose a good representation for each node in attributed networks should be generated from the information manifested in its local context, as well as be capable of recovering the local context. The subgraph centered at the target node together with associated node attributes well reflects the context. Thus recovering the context means reconstructing node attribute in local subgraph.

We directly measure the Euclidean distance between the reconstructed attribute matrix $X'$~(Output of graph context decoder) and the original input attribute matrix $X$ as loss function, which is formalized as follows:
\begin{equation}
L_c = {\left\|X - X^{\prime}\right\|}^{2}_{F}
\end{equation}
Our decoding mechanism ensures that node representations learned by this simple loss functions can recover the the attributed local subgraph.

We add L2 regularization on all the parameters of model with $\lambda$ being the hyper-parameter to control its weight. All the parameters in our framework are trained by minimize the overall loss using gradient descent. 

\subsection{Time Complexity and Speedup}
In this section, we analyse the time complexity of ANAE and show how to speed up ANAE.

\subsubsection{\textbf{Time Complexity}} 

The time complexity of graph attention layer is $O(|V|F_{in}F_{out} + |E|F_{out})$ ~\cite{velickovic2018graph}, where $F_{in}$ is the number of input features, $F_{out}$ is the number of output features, and |V| and |E| are the numbers of nodes and edges in the graph, respectively. As ANAE stacks several layers of Graph attention layer, the time complexity of ANAE is $O(|V|FD_{max} + |E|D_{max})$, where $D_{max}$ is max dimension of hidden layer, $|F|$ is the dimension of input feature. The time complexity of loss function is $O(|V|F)$. Thus the total time complexity of our model is  $O(|V|FD_{max} + |E|D_{max})$.

\subsubsection{\textbf{Speedup}} 

Graph attention layer takes the whole graph~(adjacency matrix $A$ and attribute matrix $X$) as input, and both memory and time cost for computing are related to the number of nodes $|V|$. As a result, this layer can not be directly applied to large-scale networks. Mini-batch is adopted in our implementation used to avoid inputting the whole graph, but it is still computing costly as aggregation operator depends on lots of neighbors. Specifically, neighbor nodes within $k$ hops to the target node are attached to the input in order to learn its embedding when stacking $k$ layers of graph attention networks. 
%Unfortunately, in real-world network neighbors will soon cover the whole network with increase of order $k$. 
In order to reduce the computation cost, i.e., reducing the number of neighbor nodes in a mini-batch, we follow the idea from GraphSAGE~\cite{hamilton2017inductive} and randomly sample a fixd number of neighbors for each node to update its representation in each mini-batch. Only constant number of nodes is required for each mini-batch, thus speed up our model.

\section{Experiments}\label{sec5}

We evaluate our proposed model on real-world datasets at two commonly adopted tasks, i.e., link prediction and node classification~\cite{liao2018attributed,huang2017accelerated,huang2017label,gao2018deep}. Link prediction checks the ability of nodes' representations to reconstruct network structure and predict future links, while node classification verifies whether node embeddings learned by our model are effective for node classification tasks. 
%We also compare with some supervised graph neural networks on semi-supervised node classification. Moreover, detailed analysis for the performance of our proposed model is involved in this section. 
%We conduct a detailed analysis to probe where the performance improvement comes from.
%For better demonstrating the advantages of our encoder-decoder framework, we further offer a thorough analysis comparing with one of the baselines, which also leverages the interaction between network structure and attribute information but ignore the attribute reconstruction in the decoder. 
%In addition, we also give an analysis about the effect of hyper-parameter in our model.
\subsection{Dataset}

\begin{table}[h]
\centering
\makegapedcells
\setlength{\abovecaptionskip}{0pt}%
\setlength{\belowcaptionskip}{0pt}%
\caption{Statistics of Datasets}
\begin{tabular}{ccccc}
\hline
%\toprule
\textbf{Datasets}& \textbf{Nodes}& \textbf{Edges}& \textbf{Classes}& \textbf{Features}\\
\hline
Cora& 2,708& 5,429& 7& 1,433\\
Citeseer& 3,327& 4,732& 6& 3,703\\
Wiki& 2,405& 17,981& 17& 4,973\\
Pubmed& 19,717& 44,338& 3& 500\\
\hline
%\bottomrule
\end{tabular}
\label{tb1}
\end{table}

We conduct experiments on four real-world datasets, i.e., Cora~\cite{yang2015network, liu2018content, gao2018deep}, Citeseer~\cite{yang2015network, liu2018content, gao2018deep}, Wiki~\cite{yang2015network, liu2018content, gao2018deep} and Pubmed~\cite{gao2018deep}. Cora, Citeseer and Pubmed are three citation networks where nodes are articles and edges indicate citations between articles. In these three datasets, citation relationships are viewed as undirected edges for simplicity. Attributes associated with nodes are extracted from the title and the abstract of each article, represented as sparse bag-of-word vectors. Stop words and low-frequency words are removed in preprocessing. Wiki dataset is a web page network, where nodes represent web pages and edges are hyper links among web pages. Text information on the web pages is processed in a similar way as the other three datasets to extract the attributes. Each node in the four datasets only has one label, indicating which class the node belongs to. Statistics of these datasets, including number of nodes~(Nodes), number of edges~(Edges), number of categories~(Classes) and the dimension of attributes~(Features), are summarized in Table~\ref{tb1}.
\subsection{Experiment Set-up}
\subsubsection{\textbf{Model Set-up}}
%To demonstrate the ability of our encoder-decoder framework for attributed network,
In experiments, the number of layers in graph context encoder is set to be 2. The dimensions of hidden representations in two encoder layers are set to be 128 and 64 respectively. The number of attention heads is set to be $K=8$ for the first encoder layer, and $K=1$ for the second layer. We stack two decoder layers for graph context decoder. The first decoder layer has 128 hidden units with $K=8$ attention heads. The dimension of the output of second decoder layer is set to be the dimension of input attributes and the second decoder layer has $K=1$ attention head. We also add dropout~(dropout probability$=0.5$) and L2 regularization~($\lambda=5e-4$) to prevent overfitting, and train our models using Adam with a learning rate of 0.001. Weights are all initialized by glorot~\cite{glorot2010understanding} that brings substantially faster convergence. 

 All these models are implemented in tensorflow~\cite{H-abadi2016tensorflow}, a widely used deep learning tool. We optimize all the hyper-parameters using a validation set.

\begin{table*}[!htbp]
\centering
\makegapedcells
\setlength{\abovecaptionskip}{0pt}%
\setlength{\belowcaptionskip}{0pt}%
\caption{Result of link prediction}
\begin{tabular}{c|ccccccccc}
        %\hline
        %\multicolumn{10}{c}{Datasets}& &  &  &  &  &  &  &  & \\
		\hline
		%\toprule
		\multirow{2}{*}{\textbf{Categories}}& ~ & \multicolumn{2}{c}{\textbf{Cora}} & \multicolumn{2}{c}{\textbf{Wiki}} & \multicolumn{2}{c}{\textbf{Citeseer}} & \multicolumn{2}{c}{\textbf{Pubmed}} \\
		\cline{2-10}
		%\cline
		~&\textbf{Methods} & AUC & AP & AUC & AP & AUC & AP & AUC & AP\\
		\hline
		\multirow{2}{*}{Attribute-only}&SVD	& 79.10 & 82.46 & 83.70 & 87.98 & 85.82 & 88.76 & 85.98 & 87.60\\
		&AE	& 79.14 & 79.40 & 77.26 & 82.08 & 81.39 & 82.08 & 85.15 & 84.82\\
		\hline
		\multirow{2}{*}{Structure-only}&DW 	& 80.53 & 82.79 & 81.27 & 82.39 & 73.22 & 76.21 & 76.88 & 74.73\\
		&SDNE& 77.87 & 81.95 & 81.68 & 82.66 & 74.46 & 78.91 & 77.91 & 75.29\\
		\hline
		\multirow{6}{*}{Attribute+Structure}&DW+SVD	& 81.06 & 83.13 & 89.57 & 91.10 & 73.92 & 76.72 & 76.92 & 74.76\\
		~&TADW	& \underline{93.01} & \underline{93.95} & \underline{92.19} & \underline{93.10} & \underline{94.51} & \underline{95.67}& 94.71 & 95.01\\
		~&DANE	& 88.19 & 89.56 & 91.01 & 92.45 & 84.93 & 84.68 & 87.76 & 89.69\\
		~&STNE	& 89.90 & 88.77 & 88.87 & 88.75 & 93.51 & 94.61 & 92.77 & 91.65\\
		~&DGI	&  78.53& 80.50 &  78.18&  79.30&  81.31& 82.50 &  81.66& 83.13\\
		~&GAE	& 91.47 & 92.37 & 91.81 & 92.91 & 90.52 & 91.59 & \underline{95.93} & \textbf{96.25}\\
		~&VGAE	& 91.70 & 92.64 & 91.17 & 92.49 & 90.96 & 92.97 & 94.05 & 94.42\\
		~&GAE (GAT) & 94.44 & 94.52 & 93.77 & 94.70 & 93.67 & 93.57 & 95.44 & 95.09\\
		\hline
		%\bottomrule
		%ANAE-decoder	   &94.44& 94.52 &  93.66&  93.57&  93.63&  94.59&  95.33&  94.67 \\
		%~&ANAE-attention	& 87.08 & 87.34 & 93.88 & 93.90 & 87.47 & 89.2 & 93.56 & 92.09\\
		Our Model & ANAE	& \textbf{96.70} & \textbf{97.46} & \textbf{94.54} & \textbf{95.13} & \textbf{96.99} & \textbf{96.92} & \textbf{96.91} & 96.10\\
		%& Improvement	   & 3.96\% & 3.73\% &  2.55\% & 2.18\% & 2.62\% &  1.30\% &  1.02\% &  -0.15\% \\
		\hline
	\end{tabular}
	\label{tb2}
\end{table*}

\subsubsection{\textbf{Baselines}}
We compare our model with the following baselines at both link prediction and node classification tasks. All the baselines fall into three categories, namely ``Attribute-only'', ``Structure-only'' and ``Attribute+Structure''. Models in ``Attribute-only'' group leverage node attributes information only to extract node representations, from which we select SVD and auto-encoder as our baselines. ``Structure-only'' models consider structure information only, i.e., preserving structural proximity in embedding space, while ignoring attribute information. In this group we choose Deepwalk and SDNE as our baselines. Methods in ``Attribute+Structure'' group capture both nodes attributes and structure proximity simultaneously and we consider several state-of-the-art algorithms as our baselines. A detailed description of our baselines is illustrated as follows:

%\begin{itemize}
%\item 
\textbf{Singular Value Decomposition~(SVD)}: SVD~\cite{golub1970singular} is a linear model that can extract node representations by decomposing the node attribute matrix. Following the set in~\cite{yang2015network}, we reduce the dimension of node attributes to 200 via SVD.

%\item 
\textbf{Auto-Encoder~(AE)}: AE~\cite{hinton2006reducing} is the conventional auto-encoder model with nodes attributes as input only. The number of hidden units is set the same as the ANAE.

%\item 
\textbf{DeepWalk~(DW)}: DW~\cite{perozzi2014deepwalk} learns embedding using structural information only. DeepWalk learns the node embedding from a collection of random walks using skip-gram with hierarchical softmax. As for the parameters, the number of random walks is 10, the number of vertex per walk $\gamma = 80$, window size $t = 10$ and embedding dimension $k = 128$.

%\item 
\textbf{SDNE}: SDNE~\cite{wang2016structural} is a deep model that capture both first-order and second-order proximity of nodes in embedding with only structure information being considered. The structure of hidden units in SDNE is set the same as ANAE, and the hyper-parameters are tuned by using grid search on the validation set.

%\item 
\textbf{DW+SVD}: DW+SVD concatenates representations leaned by DeepWalk and SVD.

%\item 
\textbf{TADW}: TADW~\cite{yang2015network} belongs to translation model we mentioned in Section \ref{sec3}, and it utilizes both network structure and text information to learn embedding. We set the dimension of representations to be $160$ and the coefficient of regularization term to be $0.2$ as mentioned in the paper~\cite{yang2015network}.

%\item 
\textbf{STNE}: STNE~\cite{liu2018content} is another translation model that translate attributes associate with nodes into their identities with structure information encoded in random walk path. For all the datasets, we generate 10 random walks that start at each node, and the length of the walks is set to 10. For Cora Citeseer and Wiki which are used in ~\cite{liu2018content}, we use the same architecture of model and hyper-parameters as in ~\cite{liu2018content}. The neural networks have 9 layers for Pubmed with dropout probability $p=0.2$.

%\item 
\textbf{DANE}: DANE~\cite{gao2018deep} use two independent auto-encoder to model attributes and structure information respectively with several regularizations in hidden representation. We use the same architecture and hyper-parameters of DANE as in ~\cite{gao2018deep}.

%\item 
\textbf{GAE/VGAE}: GAE and VGAE~\cite{kipf2016variational} simultaneously model network structure and node attributes using graph convolution layer and reconstruct network . Hyper-parameters are set the same as in their paper. We train the model for a maximum of 200 iterations using Adam~\cite{kingma2014adam} with a learning rate of $0.01$. 

\textbf{DGI}: DGI is an unsupervised approach for learning node representations , which relies on maximizing mutual information between patch representations and corresponding high-level summaries of graphs~\cite{velickovic2018deep}. We use the same architecture and hyper-parameters of DGI for transductive learning as in ~\cite{velickovic2018deep}.

%\item 
\textbf{GAE~(GAT)}: We replace the graph convolution layer used in GAE with graph attention layer and named this baseline GAE~(GAT). The dimensions of hidden representations in two encoder layers are set to be 128 and 64 respectively. The number of attention heads is set to be $K=8$ for the first encoder layer, and $K=1$ for the second layer. We set dropout probability $p=0.5$ and L2 regularization~($\lambda=5e-4$). We add this baseline to demonstrate that our decoder is more effective than reconstructing network structure directly.
%\end{itemize}

\subsubsection{\textbf{Task Set-up}}
After learned embedding for each node using different methods, we adopt embedding useage strategy for downstream task as many other works do~\cite{grover2016node2vec, gao2018deep, kipf2016variational, yang2015network}. For link prediction, we get link probability of node pair by calculating inner product of their embeddings. For node classification, we predict the label of each node by a logistic regression classifier~(LR) with L2 regularization using node embedding as input.

\subsection{Link Prediction}\label{sec_lp}
In this section, we evaluate the ability of learned embeddings to reconstruct network and predict future connections via link prediction. We generate the dataset as many other works do ~\cite{grover2016node2vec, wang2016structural, kipf2016variational}. We split the edges in the network according to the ratio of 85\%, 5\% and 10\% as positive instances for training, validation and testing, respectively. For test set, we add some negative samples by randomly sampling some unconnected node pairs and the ratio of positive samples and negative samples is kept as 1:1. After having obtained the embeddings for each node, we get link probability by calculating inner product of the embeddings on test data. 
We adopt the area under the ROC-curve~(AUC) and average precision from prediction scores~(AP) as the evaluation metrics. We report the results in Table \ref{tb2}, we bold the best results and underline the second best results. We summarize the following observations and analyses:
%\begin{itemize}
%\item 

``Attribute-only'', especially SVD, achieves comparable or better results than ``Structure-only'' methods in all datasets. The reason is that all these datasets are assortative networks, in which nodes with similar attributes are more likely to connect with each other.

%\item 
We also observe that ``Attribute+Structure'' methods that incorporate both node attributes and network structure information improve the link prediction performance. TADW, GAE and VGAE get better results than other ``Attribute+Structure'' methods. The superiority may result from the fact that they directly optimize to reconstruct adjacency matrix, which is highly related to the link prediction task. 
GAE~(GAT) almost consistently better than GAE except get comparable result in Pubmed. Because nodes in Pubmed have less attributes comparing with other dataset and GAT may not be well learned by reconstructing network structure.

%\item 
Our ANAE model achieves significant improvements in AUC and AP over all the baselines in all four datasets, as shown in Table \ref{tb2}. %Our model get 3.97\% AUC and 3.72\% AP gain over the best performance of all the baselines in Cora data. 
Our model well incorporates node attributes and network structure by encoding and decoding attributed local subgraph, thus achieving better results.
\begin{table*}[!htbp]
\centering
\makegapedcells
\setlength{\abovecaptionskip}{0pt}%
\setlength{\belowcaptionskip}{5pt}%
\caption{Node classification result of Cora}
\begin{tabular}{c|cccccccccc}
		\hline
		%\toprule
		\multirow{3}{*}{\textbf{Methods}}&\multicolumn{10}{c}{\% Labeled Nodes}\\
		\cline{2-11}
		 & \multicolumn{2}{c}{\textbf{10\%}} & \multicolumn{2}{c}{\textbf{20\%}} & \multicolumn{2}{c}{\textbf{30\%}} & \multicolumn{2}{c}{\textbf{40\%}} & \multicolumn{2}{c}{\textbf{50\%}} \\
		\cline{2-11}
		%\cline
		& Micro-F1 & Macro-F1 & Micro-F1 & Macro-F1 & Micro-F1 & Macro-F1 & Micro-F1 & Macro-F1 & Micro-F1 & Macro-F1\\
		\hline
		SVD	& 47.44&	34.84&	60.86&	54.60&	66.70&	62.43&	68.54&	65.20&	69.55&	66.64\\
		AE	& 68.00&	64.17&	71.38&	68.11&	71.30&	68.22&	72.36&	69.68&	73.26&	69.66\\
		\hline
		DW	& 76.53&	75.26&	78.58&	77.03&	79.86&	77.95&	80.5&	78.26&	81.33&	79.19\\
		SDNE & 76.84&	75.13&	79.64&	77.63&	79.98&	78.26&	81.31&	78.73&	82.12& 79.72\\
		\hline
		DW+SVD	& 76.65&	74.91&	80.37&	79.08&	81.63&	80.30&	83.36&	81.92&	83.97&	82.44\\
		TADW	& 77.92&	75.34&	82.04&	80.65&	83.50&	82.30&	83.89&	82.63&	84.58&	82.91\\
		DANE	& 78.01&	76.09&	80.39&	78.76&	81.86&	80.05&	82.15&	80.50&	82.50&	80.58\\
		STNE	& 81.83&	80.35&	84.31&	82.34&	84.75&	82.84&	86.27&	84.59&	86.55&	84.53\\
		DGI	& 30.25&	6.8&	30.76&	7.7&	33.12&	11.59&	36.87&	17.39&	44.22&	28.06\\
		GAE	& 80.39&	79.30&	81.31&	80.32&	82.22&	81.84&	82.27&	80.80&	82.35&	81.39\\
		VGAE	& 83.02&	81.23&	83.29&	82.36&	84.28&	83.96&	84.61&	84.35&	85.96&	84.56\\
		GAE (GAT) & \underline{83.63} & \underline{82.69} & \underline{85.14} & \underline{84.01} & \underline{85.70} & \underline{84.80} & \underline{86.76} & \underline{85.81} & \underline{86.77} & \underline{85.93}\\
		\hline
		%ANAE-decoder& 84.04&	82.92&	85.28&	84.29&	85.81&	84.51&	86.13&	85.47&	86.78&	86.09\\
		%ANAE-attention	& 78.22&	76.29&	81.12&	79.31&	82.64&	81.901&	82.15&	80.974&	83.23&	82.65\\
		ANAE	& \textbf{84.7}&	\textbf{83.73}&	\textbf{85.74}&	\textbf{85.21}&	\textbf{86.50}&	\textbf{85.77}&	\textbf{86.77}&	\textbf{85.84}&	\textbf{87.37}&	\textbf{86.29}\\
		%Improvement & 2.02\%&	3.07\%&	1.69\%&	3.46\%&	2.06\%&	2.15\%&	0.58\%&	1.47\%&	0.94\%&	2.04\%\\
		\hline
		%\bottomrule
	\end{tabular}
	\label{tb3}
\end{table*}

\begin{table*}[!htbp]
\centering
\makegapedcells
\setlength{\abovecaptionskip}{0pt}%
\setlength{\belowcaptionskip}{5pt}%
\caption{Node classification result of Wiki}
\begin{tabular}{c|cccccccccc}
		\hline
		%\toprule
		\multirow{3}{*}{\textbf{Methods}}&\multicolumn{10}{c}{\% Labeled Nodes}\\
		\cline{2-11}
		& \multicolumn{2}{c}{\textbf{10\%}} & \multicolumn{2}{c}{\textbf{20\%}} & \multicolumn{2}{c}{\textbf{30\%}} & \multicolumn{2}{c}{\textbf{40\%}} & \multicolumn{2}{c}{\textbf{50\%}} \\
		\cline{2-11}
		%\cline
		& Micro-F1 & Macro-F1 & Micro-F1 & Macro-F1 & Micro-F1 & Macro-F1 & Micro-F1 & Macro-F1 & Micro-F1 & Macro-F1\\
		\hline
		SVD	& 65.32&	49.60&	72.45&	58.08&	75.59&	61.47&	76.51&	63.14&	77.61&	65.92\\
		AE	& 53.99&	41.28&	61.69&	52.90&	65.00&	54.73&	66.39&	57.97&	68.08&	59.25\\
		\hline
		DW	& 57.95&	42.28&	62.71&	48.74&	65.34&	52.77&	65.97&	54.64&	67.15&	55.70\\
		SDNE & 53.99&	43.48&	56.79&	46.50&	59.84&	50.31&	60.84&	50.76&	61.98&	52.03\\
		\hline
		DW+SVD	& 64.24&	48.39&	71.31&	56.20&	75.14&	60.01&	75.56&	61.34&	76.97&	64.01\\
		TADW	& 67.64&	50.10&	73.04&	58.14&	76.17&	63.01&	78.03&	64.66&	79.00&	66.92\\
		DANE	& \underline{72.98}&	\underline{59.97}&	\underline{75.00}&	63.58&	77.26&	64.14&	77.52&	66.52&	78.30&	67.84\\
		STNE	& 71.31&	56.10&	74.74&	\underline{64.04}&	\underline{77.73}&	\textbf{70.62}&	\underline{79.21}&	\underline{70.04}&	\underline{80.05}&	\underline{69.47}\\
		DGI	& 25.67&11.72	&	26.70&11.94	&28.02&	12.39&	31.94&	23.92&38.58	&	22.39\\
		GAE	& 68.82&	49.75&	70.60&	53.28&	70.79&	54.24&	72.14&	57.02&	72.48&	55.95\\
		VGAE	& 65.54&	46.77&	66.73&	49.30&	68.82&	51.61&	69.43&	50.64&	68.99&	51.49\\
		GAE (GAT) & 67.25 & 49.01 & 69.28 & 52.28 & 70.01 & 52.21 & 71.31 & 54.19 & 71.57 & 54.31\\
		\hline
		%ANAE-decoder& 67.25&	48.90&	67.67&	50.14&	70.01&	51.94&	69.71&	51.73&	70.65&	52.39\\
		%ANAE-attention	& \textbf{74.96}&	63.16&	76.76&	64.99&	78.26&	68.60&	79.48&	69.09&	80.13&	71.45\\
		ANAE	& \textbf{74.82}&	\textbf{63.83}&	\textbf{77.65}&	\textbf{67.81}&	\textbf{78.44}&	69.19&	\textbf{79.62}&	\textbf{70.67}&	\textbf{80.38}&	\textbf{72.34}\\
		%Improvement & 2.52\%&	6.43\%&	3.53\%&	5.88\%&	0.91\%&	-2.02\%&	0.51\%&	0.90\%&	0.41\%&	4.13\%\\
		\hline
		%\bottomrule
	\end{tabular}
	\label{tb4}
\end{table*}

\begin{table*}[!htbp]
\centering
\makegapedcells
\setlength{\abovecaptionskip}{0pt}%
\setlength{\belowcaptionskip}{5pt}%
\caption{Node classification result of Citeseer}
\begin{tabular}{c|cccccccccc}
		\hline
		%\toprule
		\multirow{3}{*}{\textbf{Methods}}&\multicolumn{10}{c}{\% Labeled Nodes}\\
		\cline{2-11}
		& \multicolumn{2}{c}{\textbf{10\%}} & \multicolumn{2}{c}{\textbf{20\%}} & \multicolumn{2}{c}{\textbf{30\%}} & \multicolumn{2}{c}{\textbf{40\%}} & \multicolumn{2}{c}{\textbf{50\%}} \\
		\cline{2-11}
		%\cline
		 & Micro-F1 & Macro-F1 & Micro-F1 & Macro-F1 & Micro-F1 & Macro-F1 & Micro-F1 & Macro-F1 & Micro-F1 & Macro-F1\\
		\hline
		SVD	& 61.14&	53.19&	66.40&	58.56&	66.95&	59.73&	68.05&	61.31&	69.39&	62.83\\
		AE	& 68.03&	59.02&	68.99&	59.69&	69.30&	60.10&	69.89&	60.43&	70.22&	61.01\\
		\hline
		DW	& 50.15&	46.18&	54.01&	49.51&	56.24&	51.03&	56.32&	51.25&	56.83&	51.90\\
		SDNE & 53.79&	49.33&	55.39&	50.35&	57.03&	51.97&	57.41&	52.35&	58.68&	53.60\\
		\hline
		DW+SVD	& 52.84&	48.77&	57.99&	53.51&	61.37&	56.25&	64.13&	58.98&	67.29&	62.21\\
		TADW	& 67.02&	60.89&	70.08&	64.84&	71.76&	66.78&	72.19&	67.13&	72.92&	68.22\\
		DANE	& 63.64&	59.83&	67.24&	61.83&	68.69&	64.96&	72.21&	68.30&	72.30&	67.78\\
		STNE	& 66.37&	61.67&	71.45&	\underline{66.72}&	\underline{73.20}&	\underline{68.84}&	\underline{73.96}&	\underline{70.18}&	\underline{74.58}&	\underline{70.84}\\
		DGI	& 31.67&	17.22&	47.69&	35.07&	55.64&	44.33&	61.43&	50.69&	64.08&	54.06\\
		GAE	& 60.52&	52.94&	60.12&	53.13&	60.59&	52.27&	61.11&	52.87&	62.00&	53.38\\
		VGAE	& 64.91&	58.24&	66.44&	59.98&	67.06&	60.65&	67.71&	60.51&	68.25&	62.18\\
		GAE (GAT) & \underline{70.69} & \underline{63.53} & \underline{71.80} & 65.19 & 72.11 & 66.03 & 72.77 & 66.29 & 73.41 & 66.95 \\
		\hline
		%ANAE-decoder	& 69.37&	62.05&	70.9&	62.39&	70.66&	63.12&	71.18&	63.03&	71.85&	64.84\\
		%ANAE-attention	& 70.85&	65.08&	71.24&	65.90&	72.36&	66.82&	72.77&	66.50&	73.05&	67.17\\
		ANAE	& \textbf{72.61}&	\textbf{67.76}&	\textbf{74.18}&	\textbf{69.18}&	\textbf{74.33}&	\textbf{70.06}&	\textbf{75.51}&	\textbf{71.47}&	\textbf{76.04}&	\textbf{71.76}\\
		%Improvement & 8.34\%&	9.87\%&	3.82\%&	3.68\%&	1.54\%&	1.77\%&	2.09\%&	1.83\%&	1.95\%&	1.29\%\\
       \hline
       %\bottomrule
	\end{tabular}
	\label{tb5}
\end{table*}

\begin{table*}[!htbp]
\centering
\makegapedcells
\setlength{\abovecaptionskip}{0pt}%
\setlength{\belowcaptionskip}{5pt}%
\caption{Node classification result of Pubmed}
\begin{tabular}{c|cccccccccc}
		\hline
		%\toprule
		\multirow{3}{*}{\textbf{Methods}}&\multicolumn{10}{c}{\% Labeled Nodes}\\
		\cline{2-11}
		 & \multicolumn{2}{c}{\textbf{10\%}} & \multicolumn{2}{c}{\textbf{20\%}} & \multicolumn{2}{c}{\textbf{30\%}} & \multicolumn{2}{c}{\textbf{40\%}} & \multicolumn{2}{c}{\textbf{50\%}} \\
		\cline{2-11}
		%\cline
		 & Micro-F1 & Macro-F1 & Micro-F1 & Macro-F1 & Micro-F1 & Macro-F1 & Micro-F1 & Macro-F1 & Micro-F1 & Macro-F1\\
		\hline
		SVD	& 64.93&	50.21&	74.41&	68.98&	78.05&	75.47&	80.36&	79.20&	81.12&	80.31\\
		AE	& 79.34&	79.57&	80.56&	80.77&	80.85&	81.00&	81.22&	81.35&	81.54&	81.67\\
		\hline
		DW	& 79.87&	78.31&	80.65&	79.18&	80.99&	79.60&	81.20&	79.80&	81.35&	79.86\\
		SDNE & 80.17&	79.31&	81.56&	79.85&	81.34&	79.65&	81.94&	80.10&	82.12&	80.18\\
		\hline
		DW+SVD	& 79.86&	78.39&	81.35&	80.00&	82.09&	80.88&	82.63&	81.49&	83.00&	81.93\\
		TADW	& 82.86&	82.75&	83.59&	83.46&	83.83&	83.71&	84.74&	84.61&	84.78&	84.64\\
		DANE	& \underline{84.28}&	\textbf{83.97}&	\underline{85.14}&	\underline{84.87}&	\underline{85.44}&	\underline{85.17}&	\underline{85.91}&	\underline{85.54}&	\textbf{86.56}&	\underline{86.35}\\
		STNE	& 83.16&	82.32&	83.73&	82.94&	84.23&	83.43&	84.50&	83.64&	84.87&	84.05\\
		DGI	& 78.51&76.39	&82.14	&81.43	&83.20	&82.65	&	83.77&	83.30&	84.19&	83.74\\
		GAE	& 83.91&	83.27&	83.96&	83.36&	84.38&	83.77&	84.34&	83.72&	84.33&	83.72\\
		VGAE	& 82.52&	81.94&	82.63&	82.01&	83.04&	82.45&	82.99&	82.48&	83.17&	82.66\\
		GAE (GAT) & 83.28 & 82.71 & 83.75 & 83.16 & 83.85 & 83.27 & 83.94 & 83.34 & 84.22 & 83.61\\
		\hline
		%ANAE-decoder& 82.68&	82.03&	83.30&	82.68&	83.38&	82.77&	83.5&	82.75&	83.66&	82.94\\
		%ANAE-attention	& 83.32&	82.84&	84.36&	84.11&	84.43&	84.13&	84.92&	84.7&	84.9&	84.47\\
   		ANAE	& \textbf{84.71}&	83.95&	\textbf{85.49}&	\textbf{85.23}&	\textbf{85.83}&	\textbf{85.58}&	\textbf{86.36}&	\textbf{86.12}&	86.46&	\textbf{86.37}\\
   		%Improvement & 0.51\%&	-0.02\%&	0.41\%&	0.42\%&	0.45\%&	0.48\%&	0.52\%&	0.67\%&	-0.11\%&	0.02\%\\
		\hline
		%\bottomrule
	\end{tabular}
	\label{tb6}
\end{table*}

%\fi

\subsection{Node Classification}
In this section, we conduct experiments on node classification to demonstrate the effectiveness of the learned embeddings for downstream tasks. All nodes attributes and edges are observed when learning embeddings. After learning the embedding for each node, a logistic regression classifier~(LR) with L2 regularization are used to classify the nodes into different labels. We use LR package provided by sklearn~\cite{scikit-learn} with default parameters. We random sample a certain number of nodes with labels as training data and the rest as test. To conduct a comprehensive evaluation, we vary the percentage of labeled nodes in training from 10\% to 50\%. We employ Macro-F1 and Micro-F1 as the metrics to evaluate the classification result. 
%and denote as Ma-F1 and Mi-F1 respectively 
We repeat the experiments 10 times and report the mean results. All hyper-parameters used for learning embedding are set the same as previous link prediction experiment in Section \ref{sec_lp}. The classification results are shown in Table \ref{tb3},\ref{tb4},\ref{tb5},\ref{tb6} respectively, and the best results are boldfaced while the second best are underlined. From these results, we have the following observations and analysis:

%\begin{itemize}
%\item 
``Structure-only'' methods outperform ``Attribute-only'' methods in Cora, get comparable result in Pubmed, while perform worse on the other two datasets. Characteristics of datasets are the main reason to explain this phenomenon. Node attributes in Wiki network contribute more on the classification of nodes as the hyperlink relationship is loose. Web pages belonging to different categories still have a high probability to have hyperlinks. Documents in both Citeseer and Wiki have more words than Cora, so ``Attribute-only'' methods perform better. Cora network has high edge density and less words in documents thus ``Structure-only'' methods get better result in this dataset.

%\item
Well-designed attributed network embedding methods~(TADW, DANE, STNE) perform better than both ``Attribute-only'' methods and ``Structure-only'' methods, because these two kinds of information describe different aspects of the same node and provide complementary information. Unfortunately, simply concatenate these two kinds of information may not improve the performance, as ``SVD+DW'' gets worse node classification result than SVD in Citeseer and Wiki dataset.
%It mainly because that these two embeddings may not have the same scalar, some dimension of embedding dominates and these two kinds of information are not well complemented by each other. 
It demonstrates that simple concatenation is not sufficient to capture the interaction between these two types of information. GAE and VGAE get poor results in Wiki, which can be explained by the design of these two models. As mentioned before, wiki network has more casual connections than other networks and the aim of GAE and VGAE is to reconstruct the observed edges, thus more likely overfitting the observed edges and bringing in much noises. We find that GAE~(GAT) is superior than GAE in Cora and Citeseer, but get worse in Wiki and Pubmed. In Wiki, reconstructing the noise edges makes GAE~(GAT) fail to capture the similarity between node pair and get worse result. In Pubmed, due to insufficient node attributes graph attention layer may not be well learned. 
%From this we believe that directly replace GCN in GAE with GAT is not a good model.
%\item

Our model outperforms all compared baselines in Cora, Citeseer and Pubmed. In Wiki network, our model achieves comparable result with STNE but still outperforms other baselines. Thus, we can conclude that our framework learns the representation of each node by encoding and decoding the attributed local subgraph is more effective than previous attributed network embedding methods.

%Our model captures the attribute information and structure information in a unified way and adopts attention mechanism to enhance proximity modeling, thus better utilizing these two complementary type of information.
%\item
We also find another interesting phenomenon that our model ANAE outperforms all compared baselines when fewer labeled nodes are available in training. As we can see in the Table,\ref{tb3},\ref{tb4},\ref{tb5},\ref{tb6} the result of almost all the baselines~(except GAE and VGAE) drop quickly when fewer labeled nodes are used in training. The reason is that these baselines do not well use neighboring nodes to preserve proximities in embedding space. By leveraging our graph context encoder, our model gets smooth representations of nodes with their neighbors', thus obtaining better result especially when lacking of label information. Although edges in Wiki are not reliable, our model still get better result in Wiki data. This is because our model better leverage node attributes---learn more accurate aggregation weights through attention mechanism and refine useful information with graph context decoder. 
%\end{itemize}

\subsection{Semi-supervised Node Classification}

\begin{table}[h]
\centering
\makegapedcells
\setlength{\abovecaptionskip}{0pt}%
\setlength{\belowcaptionskip}{0pt}%
\caption{Semi-supervised node classification results}
\begin{tabular}{ccccc}
		\hline
		%\toprule
		%\multicolumn{2}{c|}{\multirow{2}{*}{\textbf{Methods}}}&\multicolumn{3}{c}{Datasets}\\
		%\cline{3-5}
		%\cline
		 {\textbf{Labels}} & \textbf{Methods}&  \textbf{Cora} & \textbf{Citeseer} & {\textbf{Pubmed}} \\
		\hline
		\multirow{4}{*}{\textbf{N}} & Raw Features & 55.1 & 46.5 & 71.4\\
		&DW\cite{perozzi2014deepwalk} & 67.2 & 43.2& 65.3\\
		&DW+Features\cite{velickovic2018deep} & 70.7 & 51.4& 74.3\\
		&DGI\cite{velickovic2018deep} & 82.3 &71.8 & 76.8 \\
		\hline
		\multirow{5}{*}{\textbf{Y}} & LP\cite{zhu2003semi}	& 68.0 & 45.3& 63.0\\
		&Planetoid\cite{yang2016revisiting} & 75.7& 64.7& 77.2\\
		&Chebyshev\cite{defferrard2016convolutional} & 81.2 & 69.8 & 74.4 \\
		&GCN\cite{kipf2017semi}& 81.5 & 70.3 & 79.0\\
		&GAT\cite{velickovic2018graph}& 83.0 & \textbf{72.5} & 79.0\\
		\hline 
		\textbf{N}&\textbf{ANAE}& \textbf{83.2}& \textbf{72.5}& \textbf{79.7}\\
		\hline
		%\bottomrule
	\end{tabular}
	\label{semi}
\end{table}
In this section, we compare our model with semi-supervised node classification methods with state-of-the-art performance. For fairness, we use the same datasets and experimental setting as~\cite{kipf2017semi}.
We reuse the metrics already reported in (Velickovic et al.)~\cite{velickovic2018graph} for the performance of DeepWalk~(DW)\cite{perozzi2014deepwalk} , Label Propagation (LP)~\cite{zhu2003semi}, Planetoid~\cite{yang2016revisiting}, Chebyshev\cite{defferrard2016convolutional}, GCN~\cite{kipf2017semi} and GAT\cite{velickovic2018graph}. 
%DGI is an unsupervised node representation learning framework using graph convolution layer as encoder and max mutual information as decoder. We reuse the its results report in ~\cite{velickovic2018deep}. 
``DW+Features'' means concatenate attributes with embedding learned by deepwalk as representation for each node. 

All the baselines can be categorized into two groups, according to whether incorporating the label information during representations learning.  Models utilizing label information for representation learning is denoted by ``Y" in the first column in Table ~\ref{semi}. Other models that not leverage label information are denoted by ``N", and a logistic regression classifier~(LR) with L2 regularization are used to classify the nodes learned by these methods. All the results are shown in Table \ref{semi}. 

Particularly, we note that our approach gets mean classification accuracy $83.2\%$, $72.5\%$, $79.7\%$ on Cora, Citeseer and Pubmed respectively, which outperforms all the unsupervised baselines and are comparable with state-of-the-art method GAT~\cite{velickovic2018graph}.

%As show in Figure 2, our models outperform TADW in all the datasets. It demonstrates that coupling network structure with node attributes when learning embeddings are also effective for node classification task.
 %It give evidence of coupling network structure with node attributes when learning embedding improve the embedding for downstream tasks. 
 %Macro-f1 of our models gain 5.7\% and 6.3\% over GAE and VGAE respectively in Cora, 16.8\% and 26.3\% in Citeseer, and slightly better in Pubmed, which demonstrates that attribute decoder significantly improves embedding. 

\section{Analysis of ANAE}\label{sec6}

\begin{table}[h]
\centering
\makegapedcells
\setlength{\abovecaptionskip}{0pt}%
\setlength{\belowcaptionskip}{0pt}%
\caption{Link prediction results of ANAE and its variants}
\begin{tabular}{c|cccc}
		\hline
		%\toprule
		\multirow{2}{*}{\textbf{Methods}} &\multicolumn{4}{c}{Datasets}\\
		\cline{2-5}
		%\cline
		 & \textbf{Cora} & \textbf{Wiki} & \textbf{Citeseer} & {\textbf{Pubmed}} \\
		\hline
		ANAE	& \textbf{96.70} & \textbf{94.54} & \textbf{96.99}  & \textbf{96.91}\\
		\hline
		%GAE 		&91.47& 91.81& 90.52& 95.93 \\
		%ANAE-S	    &94.44 &  93.66&  93.63&   95.33 \\
		ANAE-GCN	& 87.08  & 93.88 & 87.47  & 93.56 \\
		ANAE-MLP 	&89.20 & 88.60 & 91.81& 91.49 \\
		
		\hline
		%\bottomrule
	\end{tabular}
	\label{tb-model-lp}
\end{table}

\begin{table*}[h]
%\begin{table*}
\centering
\makegapedcells
\setlength{\abovecaptionskip}{0pt}%
\setlength{\belowcaptionskip}{0pt}%
\caption{Node classification results of ANAE and its variants}
\begin{tabular}{c|ccc|ccc|ccc|ccc}
		\hline
		%\toprule
		\multirow{3}{*}{\textbf{Methods}} &\multicolumn{12}{c}{Datasets}\\
		\cline{2-13}
		%\cline
		 &\multicolumn{3}{c|}{\textbf{Cora}} & \multicolumn{3}{c|}{\textbf{Wiki}} & \multicolumn{3}{c|}{\textbf{Citeseer}} & \multicolumn{3}{c}{\textbf{Pubmed}} \\
		 \cline{2-13}
		 &10\%&30\%&50\% & 10\%&30\%&50\% &10\%&30\%&50\% & 10\%&30\%&50\% \\
		\hline
		ANAE	   & \textbf{84.70}& \textbf{86.50}& \textbf{87.37}& \textbf{74.82}& \textbf{78.44}& \textbf{80.38}& \textbf{72.61}& \textbf{74.33}& \textbf{76.04}& \textbf{84.71}& \textbf{85.83}& \textbf{86.46} \\
		\hline
		%GAE        & 80.39& 82.22& 82.35& 68.82& 70.60& 72.48& 60.52& 60.59& 62.00& 83.91& 84.38& 84.33 \\
		%ANAE-S	   & 83.60& 85.70& 86.70& 66.97& 69.06& 70.04& 70.08& 71.30& 72.10& 82.69& 83.70& 84.60 \\
		ANAE-GCN   & 78.22& 82.64& 83.23& 73.96& 78.26& 80.13& 70.85& 72.36& 73.05& 83.32& 84.43& 84.90 \\
		ANAE-MLP   & 69.64& 76.21& 79.13& 74.13& 77.02& 78.30& 70.18& 70.84& 71.18& 71.29& 76.89& 78.66 \\
		\hline
		%\bottomrule
	\end{tabular}
	\label{tb-model-nc}
\end{table*}

In this section, we give a detailed analysis of our proposed ANAE. Firstly, we verify the effectiveness of attention mechanism and GAT based graph context decoder through experiments. Then, we show the effects of choosing different latent dimensions available to our model.
\subsection{ANAE and Its Variants}\label{sec6.1}
In this section we compare ANAE with its variants ``ANAE-GCN'' and ``ANAE-MLP'' to verify the effectiveness of attention based aggregation weighting mechanism and graph context decoder respectively. 

%\subsubsection{\textbf{ANAE-GCN}}
\textbf{ANAE-GCN}
To investigate the influence of attention mechanism, we replace the attention based aggregation weighting mechanism with edge based aggregation weighting mechanism, where $\alpha_{ij} = \frac{e_{ij}}{\sum_{j\in \mathcal{N}_i}{e_{ij}}}$, $e_{ij}$ is the edge weight, and keep the framework the same as ANAE. We mark this variant as ``ANAE-GCN''. 

%\subsubsection{\textbf{ANAE-S and ANAE-MLP}}
\textbf{ANAE-MLP}
We add another variants of our models named ``ANAE-MLP'' to analyze the influence of GAT used in graph context decoder. 
%The first one inspects effectiveness of content reconstruction loss by removing graph context decoder and adopts structure reconstruction based objective function, the loss function is formalized as follows:
%\begin{equation}
%L = {\left\|A - \sigma(ZZ^T) \right\|}^{2}_{F}
%\end{equation}
%\begin{equation}
% L  = - \sum_{<u,v> \in E}\log \left( \sigma \left( \mathbf { z } _ { u } ^ { \top } \mathbf { z } _ { v } \right) \right) - Q \cdot \mathbb { E } _ { v _ { n } \sim P _ { n } ( v ) } \log \left( \sigma \left( - \mathbf { z } _ { u } ^ { \top } \mathbf { z } _ { v _ { n } } \right) \right),
%\end{equation}
% where $Q$ is the number of negative samples and we set $Q=10$. $ P _ { n } ( v )$ is the distribution to sample nodes from, uniform distribution is used in our experiments.

We replace decoder layer with two fully connected layers in ``ANAE-MLP'' in order to verify the effectiveness of graph attention layers in decoder of ANAE.
 %Under this circumstance, hidden representations of nodes are not used to reconstruction their attributed local subgraph anymore. Thus, we can verify effectiveness of our graph context decoder.

%\subsubsection{\textbf{Experiments}}
\textbf{Experiments}
%In this section we report the results of ANAE and its ablations on node classification and link prediction.
Experiments setup is the same as Section \ref{sec5} and hyper parameters of ablations, e.g., number of hidden units, are set the same as ANAE for fair comparison.  We display the AUC results for link prediction in Table \ref{tb-model-lp}, and micro-F1 results of node classification in Table \ref{tb-model-nc}. 

As shown in Table \ref{tb-model-lp} and Table \ref{tb-model-nc}, ANAE performs consistently better than ``ANAE-GCN'' in both link prediction and node classification. This phenomenon can be explained by following two reasons. First, edges in the four datasets do not contain rich information and only indicate whether two nodes are connected with each other. Thus ``ANAE-GCN'' aggregates all neighboring nodes without distinguishing importance of nodes, which limits the capacity of model. Second, attention based mechanism provides a more flexible way to capture the proximities in node attributes and network structure by reweighting the importance of neighbors.

%In all downstream tasks, we find that ANAE outperforms ``ANAE-S'' especially when edges are not reliable, for example ANAE outperforms ``ANAE-S'' with a huge margin on Wiki dataset in Table \ref{tb-model-nc}. The results may come from that ``ANAE-S'' over-emphasize local proximit as its structure reconstruction based loss have similar effect with its encoder--model the proximity of nodes. While our model aims to reconstruct local atttributed subgraph well preserved key context information of each node.

We also observe that ANAE consistently better than ``ANAE-MLP''. ANAE implicitly constrain the representation of each node to reconstruct its attributed local subgraph, because hidden representation of target node is diffused to nodes in its local subgraph and help reconstruct their attributes. But representations learned by ``ANAE-MLP'' only concentrate on reconstructing the attributes of target nodes, which may overfit some noises.

\subsection{Influence of Dimension of Hidden Representation}

\begin{figure}[!t]
\centering
\subfloat[Cora]{\includegraphics[width=0.25\textwidth]{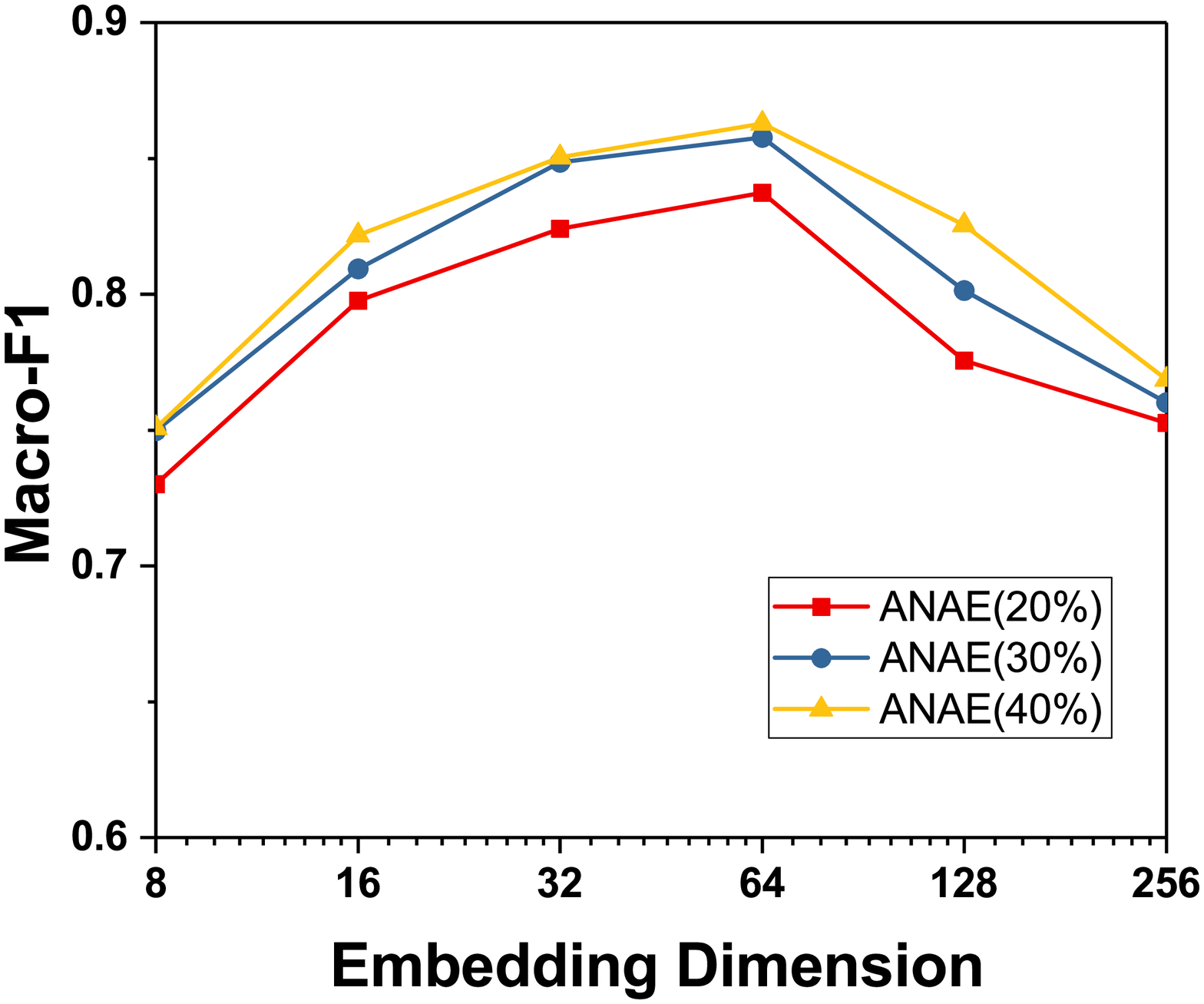}%
\label{hp_pict_1}}
\subfloat[Citeseer]{\includegraphics[width=0.25\textwidth]{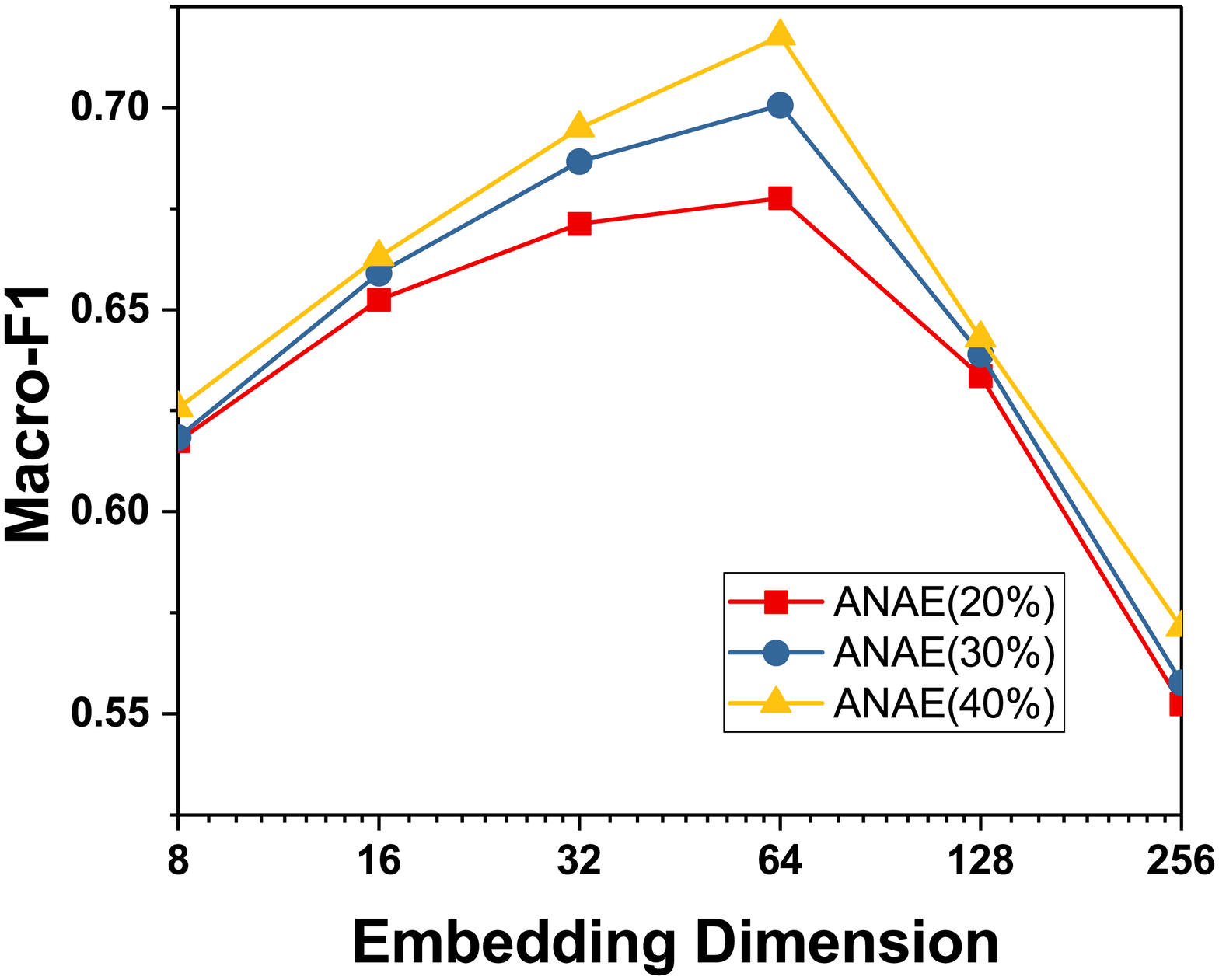}%
\label{hp_pict_2}}
\hfil
\subfloat[Wiki]{\includegraphics[width=0.25\textwidth]{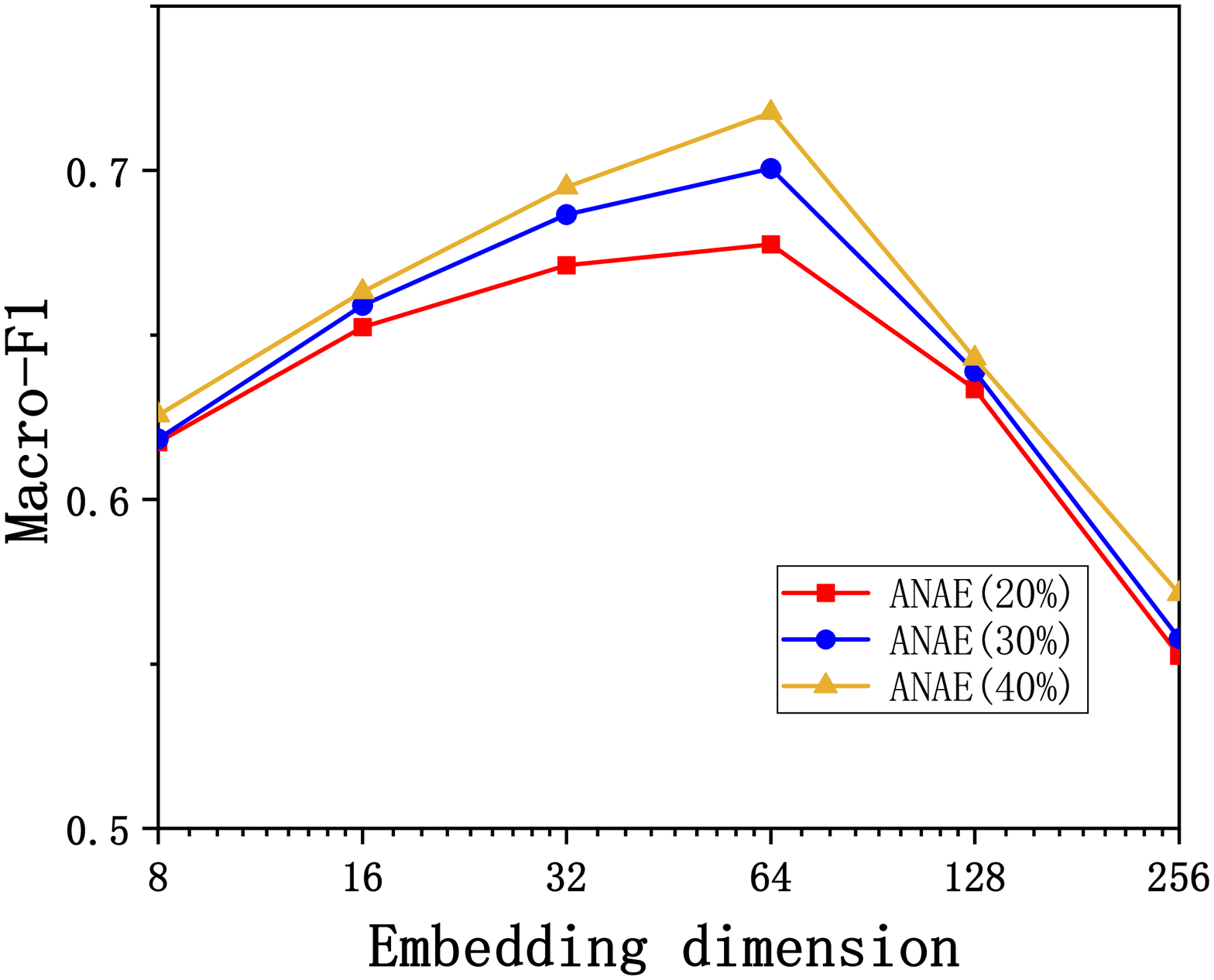}%
\label{hp_pict_3}}
\subfloat[Pubmed]{\includegraphics[width=0.25\textwidth]{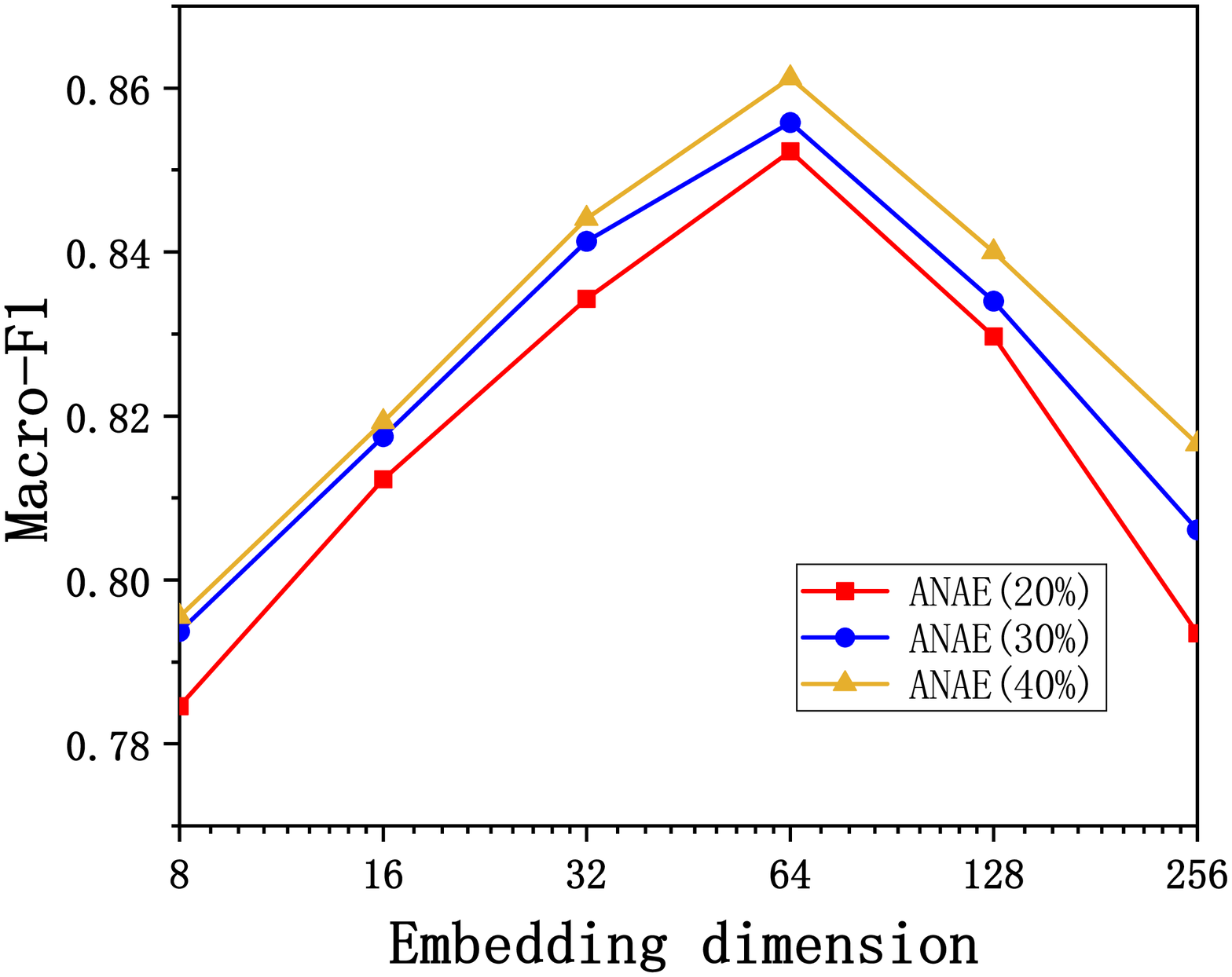}%
\label{hp_pict_4}}
\caption{Effect of different hidden dimensions.}
\label{hp_pict}
\end{figure}

Dimension of embedding is an important parameter, thus we examine how the different dimensions of embedding affect the performance of downstream tasks. We only display the result of node classification on four datasets and we get similar result in link prediction task. We vary dimension of embedding  from [8, 16, 32, 64, 128, 256], the number of units in first hidden layer is twice than the dimension of embedding. Other hyper-parameters are kept the same as mentioned in Section \ref{sec5}. Results of effect of different hidden dimensions are shown in Figure \ref{hp_pict}.

\subsection{Embedding Visualization}
\begin{figure}[!t]
\centering
\subfloat[DeepWalk]{\includegraphics[width=0.25\textwidth]{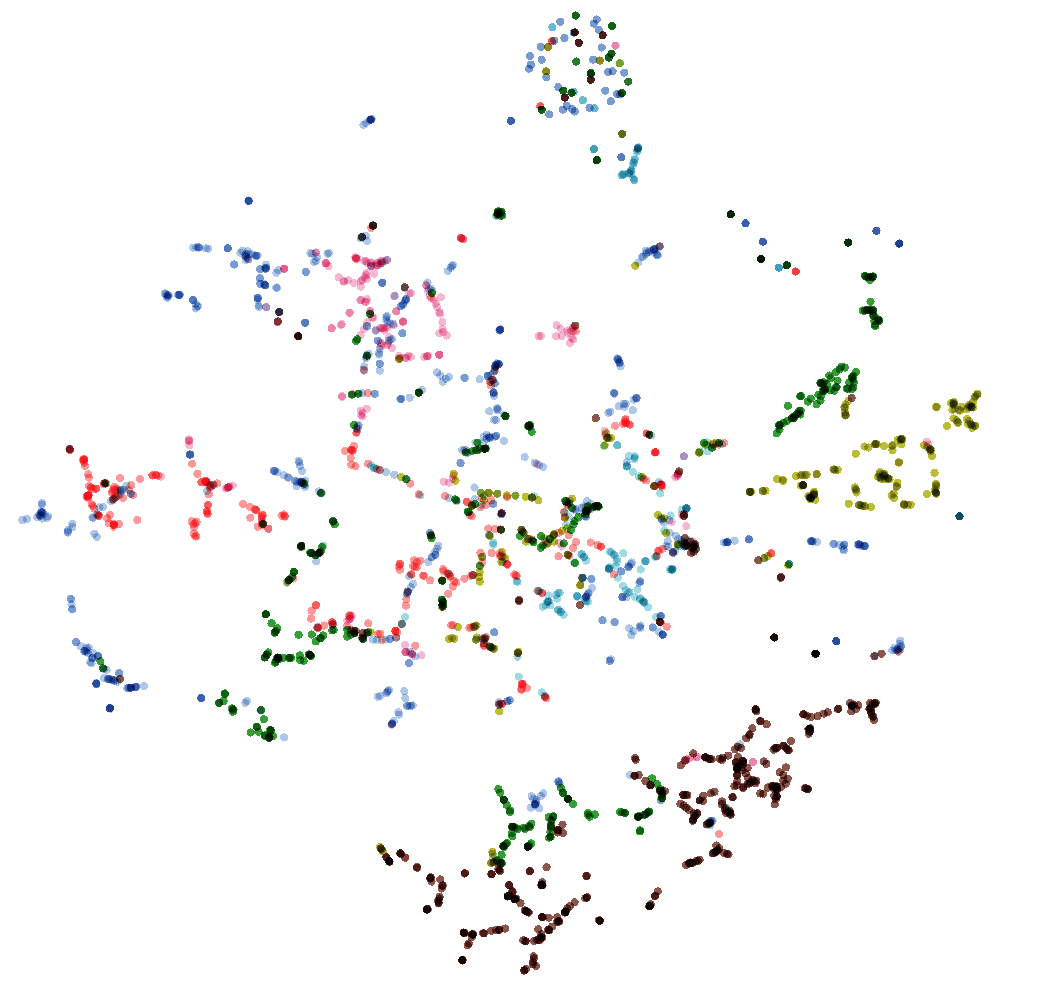}%
\label{vis_cora_1}}
\subfloat[TADW]{\includegraphics[width=0.25\textwidth]{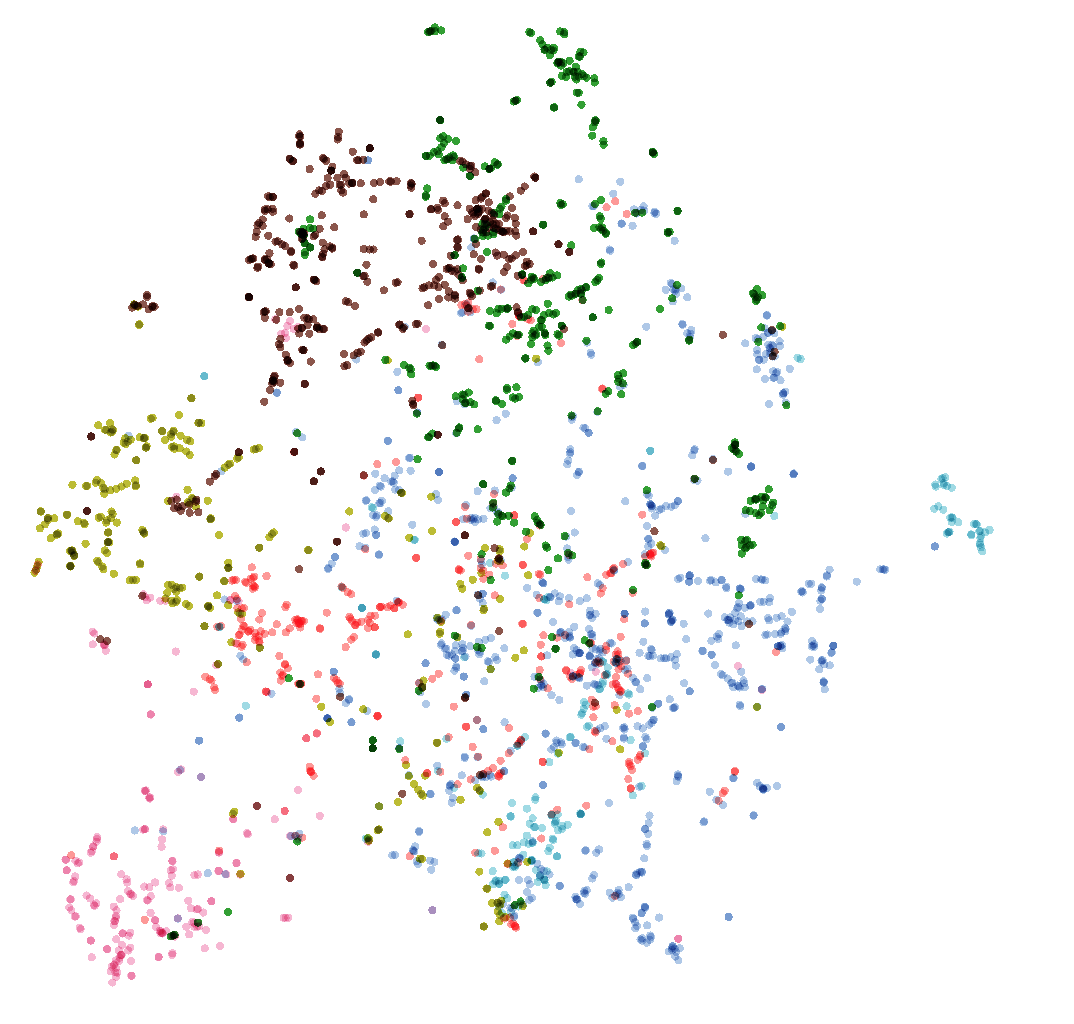}%
\label{vis_cora_2}}
\hfil
\subfloat[GAE]{\includegraphics[width=0.25\textwidth]{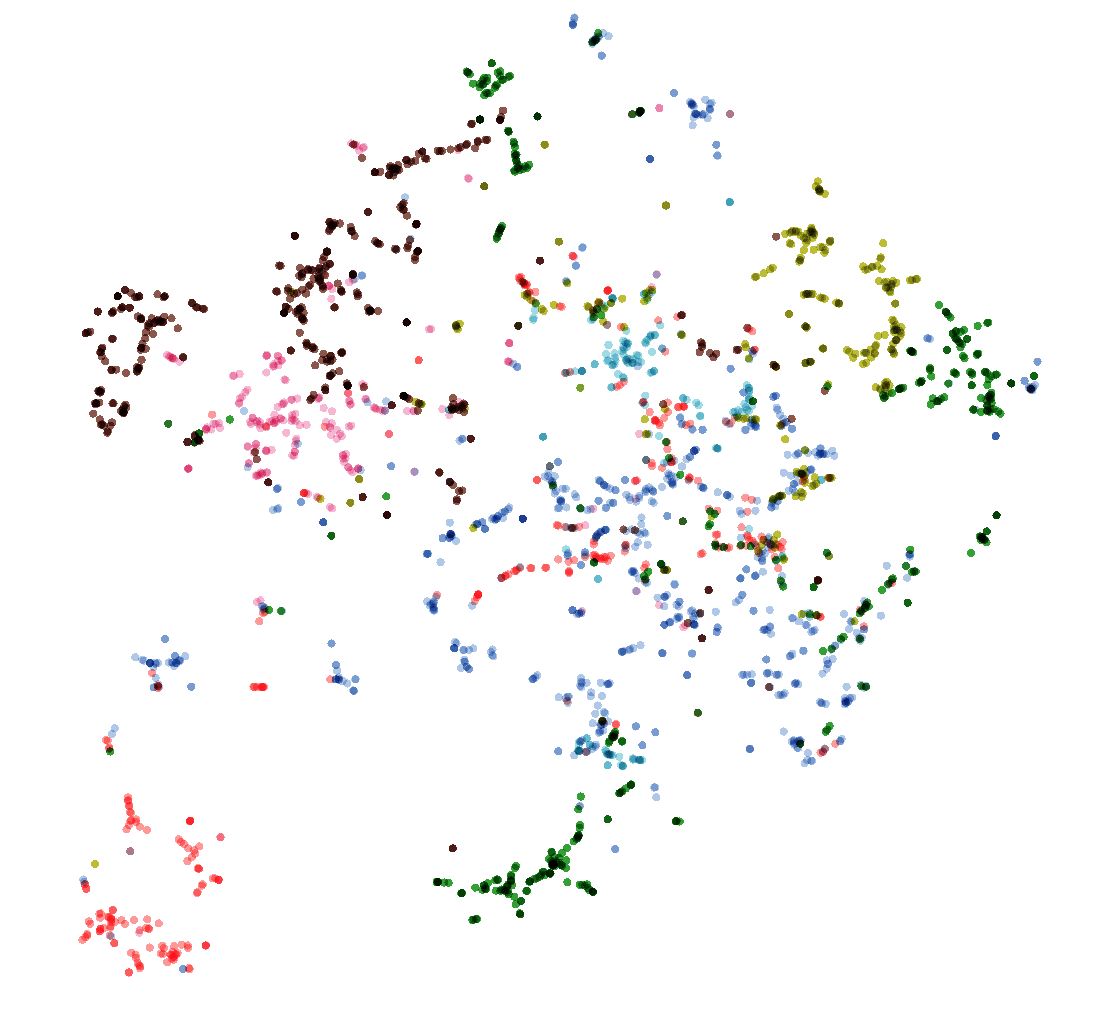}%
\label{vis_cora_3}}
\subfloat[ANAE]{\includegraphics[width=0.25\textwidth]{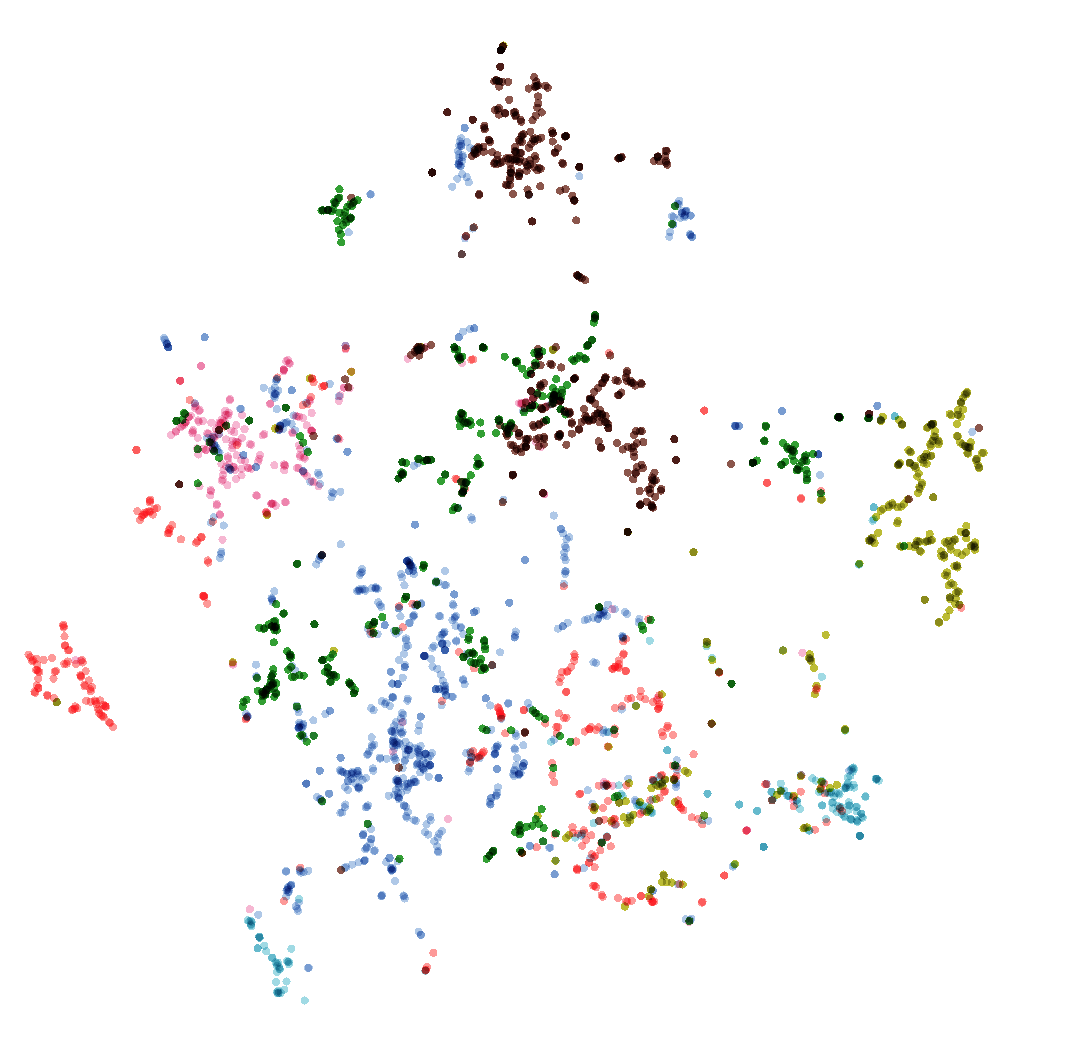}%
\label{vis_cora_4}}
\caption{Embedding Visualization on Cora. Different node colors indicate different node labels.}
\label{visual_1}
\end{figure}

\begin{figure}[!t]
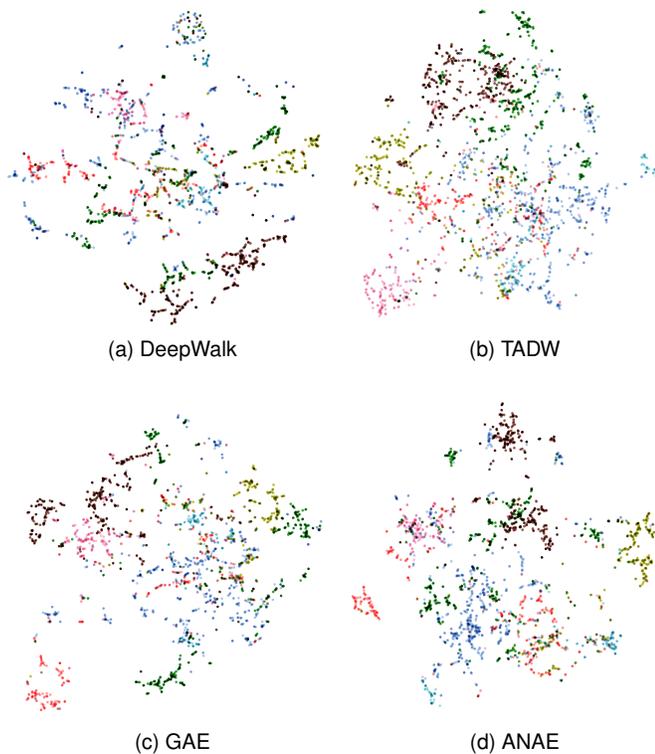

\centering
\subfloat[DeepWalk]{\includegraphics[width=0.25\textwidth]{pict/visual/cora_deepwalk.png}%
\label{vis_citeseer_1}}
\subfloat[TADW]{\includegraphics[width=0.25\textwidth]{pict/visual/cora_tadw.png}%
\label{vis_citeseer_2}}
\hfil
\subfloat[GAE]{\includegraphics[width=0.25\textwidth]{pict/visual/cora_gae.png}%
\label{vis_citeseer_3}}
\subfloat[ANAE]{\includegraphics[width=0.25\textwidth]{pict/visual/cora_anae.png}%
\label{vis_citeseer_4}}
\caption{Embedding Visualization on Citeseer. Different node colors indicate different node labels.}
\label{visual_2}
\end{figure}

In this section, we visualize the embeddings for examine the network representations intuitively. We use t-SNE toolkit~\cite{maaten2008visualizing} for embeddings visualization and compare ANAE with baselines on Cora and Citeseer. The result of Cora and Citeseer are shown in Figure~\ref{visual_1} and Figure~\ref{visual_2} respectively, each dot represents a node and different colors indicate different labels.

\section{Conclusion}\label{sec7}
In this paper, we address attributed network embedding from a novel perspective, i.e., learning node context representation for each node, and define the node context as the subgraph centered at the target node together with associated node attributes. We propose a novel attributed network auto-encoder framework, namely ANAE. For each node, graph context encoder aggregates all attribute information in its local subgraph, while graph context decoder refines its aggregated representation by reconstructing the attributes of nodes in its local subgraph. Our model captures the network structure and node attributes information in attributed local subgraph, thus has high capacity to learn good node representations for attributed network. Experimental results show that our model consistently outperforms all the benchmark algorithms in two down-stream tasks. In the future, we will explore richer semantic information in node context, e.g., exploring node polysemy.

\ifCLASSOPTIONcompsoc
  \section*{Acknowledgments}
\else
  \section*{Acknowledgment}
\fi

This work was funded by the National Natural Science Foundation of China under grant numbers 91746301 and  61425016. Huawei Shen is also funded by Beijing Academy of Artificial Intelligence (BAAI) and K.C. Wang Education Foundation.

\ifCLASSOPTIONcaptionsoff
  \newpage
\fi



%\bibliographystyle{IEEEtran}
%\begin{thebibliography}{1}


%\end{thebibliography}
\bibliographystyle{IEEEtran}
\bibliography{enhence_NE}

\vspace{-100pt}
\begin{IEEEbiographynophoto}{Keting Cen}
is currently a PhD student in the CAS Key Laboratory of Network Data Science and Technology, Institute of Computing Technology, Chinese Academy of Sciences (CAS). He received his Bachelor Degree from Harbin Institute of Technology in 2016. His research interests include network embedding, semi-supervised learning on graph and graph convolutional networks.
\end{IEEEbiographynophoto}
\vskip -2\baselineskip plus -1fil
\begin{IEEEbiographynophoto}{Huawei Shen}
is a professor in the Institute of Computing Technology, Chinese Academy of Sciences (CAS) and the University of Chinese Academy of Sciences. He received his PhD degree from the Institute of Computing Technology in 2010. His major research interests include network science, social media analytics and recommendation. He has published more than 80 papers in prestigious journals and top international conferences, including in Science, PNAS, Physical Review E, WWW, AAAI, IJCAI, SIGIR, CIKM, and WSDM. He is an Outstanding Member of the Association of Innovation Promotion for Youth of CAS. He received the Top 100 Doctoral Thesis Award of CAS in 2011 and the Grand Scholarship of the President of CAS in 2010.
\end{IEEEbiographynophoto}
\vskip -2\baselineskip plus -1fil

%\newpage
\begin{IEEEbiographynophoto}{Jinhua Gao}
is an assistant professor at Institute of Computing Technology, Chinese Academy of Sciences (CAS). He received his Ph.D. degree from Institute of Computing Technology, CAS in 2018. His major research interests include social computing and public opinion mining.
\end{IEEEbiographynophoto}
\vskip -2\baselineskip plus -1fil

\begin{IEEEbiographynophoto}{Qi Cao}
is currently a PhD student in the CAS Key Laboratory of Network Data Science and Technology, Institute of Computing Technology, Chinese Academy of Sciences (CAS). She received her Bachelor Degree from Renmin University of China in 2015. Her research interests include social media computing, influence modeling and information diffusion prediction.
\end{IEEEbiographynophoto}
\vskip -2\baselineskip plus -1fil

\begin{IEEEbiographynophoto}{Bingbing Xu}
is currently a PhD student in the CAS Key Laboratory of Network Data Science and Technology, Institute of Computing Technology, Chinese Academy of Sciences (CAS). She received her Bachelor Degree from Harbin Institute of Technology in 2016. Her research interests include geometric deep learning, graph-based data mining and graph convolution.
\end{IEEEbiographynophoto}
\vskip -2\baselineskip plus -1fil

\begin{IEEEbiographynophoto}{Xueqi Cheng}
is a professor in the Institute of Computing Technology, Chinese Academy of Sciences (ICT-CAS) and the University of Chinese Academy of Sciences, and the director of the CAS Key Laboratory of Network Data Science and Technology. His main research interests include network science, web search and data mining, big data processing and distributed computing architecture. He has published more than 200 publications in prestigious journals and conferences, including IEEE Transactions on Information Theory, IEEE Transactions on Knowledge and Data Engineering, Journal of  Statistical Mechanics, Physical Review E., ACM SIGIR, WWW, ACM CIKM, WSDM, AAAI, IJCAI, ICDM, and so on. He has won the Best Paper Award in CIKM (2011), the Best Student Paper Award in SIGIR (2012), and the Best Paper Award Runner up of CIKM (2017). He is currently serving on the editorial board of the Journal of Computer Science and Technology, the Journal of Computer, and so on. He received the China Youth Science and Technology Award (2011), the Young Scientist Award of Chinese Academy of Sciences (2010), the Second prize for the National Science and Technology Progress (2012). He is a member of the IEEE.
\end{IEEEbiographynophoto}

\end{document}